\documentclass[aps,prl,twocolumn,superscriptaddress,reprint,amsmath,amssymb]{revtex4-2}
\usepackage{graphicx}% Include figure files
\usepackage{dcolumn}% Align table columns on decimal point
\usepackage{bm}% bold math
\usepackage{latexsym}
\usepackage{array}
\usepackage{setspace}

\usepackage{multirow}

\usepackage{xcolor}
\usepackage[normalem]{ulem}

	\usepackage{amsthm}
	\theoremstyle{theorem}
	
    \usepackage[justification=centerlast]{caption}
    \usepackage[labelformat=simple]{subcaption}

    \newcommand{\beginsupplement}{%
    	\setcounter{equation}{0}
    	\renewcommand{\theequation}{S\arabic{equation}}%
    	\setcounter{table}{0}%
    	\renewcommand{\thetable}{S\arabic{table}}%
    	\setcounter{figure}{0}%
    	\renewcommand{\thefigure}{S\arabic{figure}}%
    }

\begin{document}

%\preprint{Physical Review Letter}

\title{
{Concurrence Percolation in Quantum Networks}
}% Force line breaks with \\

\author{Xiangyi~Meng}%
\affiliation{Center for Polymer Studies and Department of Physics, Boston University, Boston, Massachusetts 02215, USA}%
\affiliation{Center for Complex Network Research and Department of Physics, Northeastern University, Boston, Massachusetts 02115, USA}
\author{Jianxi~Gao}%
\email{jianxi.gao@gmail.com}
\affiliation{Department of Computer Science, Rensselaer Polytechnic Institute, Troy, New York 12180, USA}
\author{Shlomo~Havlin}
\email{havlins@gmail.com}
\affiliation{Department of Physics, Bar-Ilan University, 52900 Ramat-Gan, Israel}%
\affiliation{Center for Polymer Studies and Department of Physics, Boston University, Boston, Massachusetts 02215, USA}%

\date{\today}


\begin{abstract}
Establishing long-distance quantum entanglement, i.e.,~entanglement transmission, in quantum networks (QN) is a key and timely challenge for developing efficient quantum communication. Traditional comprehension based on classical percolation assumes a necessary condition for successful entanglement transmission between any two infinitely distant nodes: they must be connected by at least a path of perfectly entangled states (singlets). Here, we relax this condition by explicitly showing that one can focus not on optimally converting singlets but on establishing concurrence---a key measure of bipartite entanglement. We thereby introduce a new statistical theory, concurrence percolation theory (ConPT), remotely analogous to classical percolation but fundamentally different, built by generalizing bond percolation in terms of ``sponge-crossing'' paths instead of clusters. Inspired by resistance network analysis, we determine the path connectivity by series and parallel rules and approximate higher-order rules via star-mesh transforms. Interestingly, we find that the entanglement transmission threshold predicted by ConPT is lower than the known classical-percolation-based results and is readily achievable on any series-parallel networks such as the Bethe lattice. ConPT promotes our understanding of how well quantum communication can be further systematically improved versus classical statistical predictions under the limitation of QN locality---a ``quantum advantage'' that is more general and efficient than expected. ConPT also shows a percolationlike universal critical behavior derived by finite-size analysis on the Bethe lattice and regular two-dimensional lattices, offering new perspectives for a theory of criticality in entanglement statistics.
\end{abstract}


\maketitle

%\hyphenpenalty=3000 %8000
%\tolerance=8000    %3000

Recently, much attention has been given to \emph{quantum network} (QN)~\cite{QEP_acl07} (sometimes also referred to as the quantum Internet~\cite{q-internet_k08,*q-netw-summ_bfd19}) for better understanding of \emph{entanglement transmission}, i.e.,~establishing long-distance entanglement between arbitrary two nodes, as a quantum information flow from the perspective of network science~\cite{stat-of-netw_ab02,*struct-dyn-netw}.
Only local operations and classical communication, a.k.a.~LOCC~\cite{nielsen_n99} are allowed between different nodes in a QN---a limitation by locality.
In this Letter, we focus on a minimal version of QN~\cite{QEP_acl07} 
that is an $n$-node network, denoted $\mathcal{G}_{\theta}(n)$. Each link $i$ is an identical pure state $\left|\psi_i(\theta)\right\rangle=\cos\theta\left|00\right\rangle+\sin\theta\left|11\right\rangle$ weighted by the sole parameter $0\le\theta\le\pi/4$ that admits a probability measure $p:=2\sin^2\theta$ known as the optimal probability to convert $\left|\psi_i(\theta)\right\rangle$ to a \emph{singlet} (i.e.,~a maximally entangled state by $\theta=\pi/4$). Hence, a {mapping} between entanglement transmission in infinite QN and classical bond percolation theory, called classical entanglement percolation (CEP) has been discovered~\cite{QEP_acl07}. This indicates the existence of a nontrivial threshold---in terms of $p$ per link---for establishing sufficient entanglement between arbitrary two nodes. Interestingly, a scheme called quantum entanglement percolation (QEP)~\cite{QEP_acl07} shows that there are scalable quantum strategies that can change the whole network topology and thus may lower the classical percolation threshold, suggesting a ``quantum advantage'' vs.~CEP for specific network topologies. %in a much larger scenario in terms of QN. 
Generalizations to mixed states~\cite{QEP-mix-state_bdj09,*QEP-mix-state_bdj10}, tripartite entanglements~(GHZ states)~\cite{QEP-GHZ_pclla10}, and random
networks~\cite{QEP-q-swap_cc09,QEP-complex-netw_plac10,QEP-complex-netw_cc11,*QEP-complex-netw_wz11} have since been studied under the QEP scheme for further efforts on lowering the threshold, in hope of exploiting more advantage until reaching some presumed minimum threshold~\cite{QEP-GHZ_pclla10}.

{Still, all aforementioned schemes are based on the classical percolation framework. Thus, no matter how designed, the schemes have always demanded one condition to achieve entanglement transmission in infinite QN: \emph{two infinitely distant nodes must be connected by at least one path of singlets}, so that by applying a specific LOCC called ``swapping''~\cite{entangle-swap_zzhe93,*QEP-series-rule_bvk99} at in-between nodes, a singlet can eventually be established between the pair of nodes~\cite{QEP_acl07}. Naturally, a fundamental question whether \emph{in general} such a condition can be relaxed arises~\cite{QEP-detail_pcalw08,QEP-GHZ_pclla10}.} The inability of answering this within the classical percolation framework {(since the question is pertinent to the mapping itself)} substantially prevented us from a true comprehension of the quantum advantage possessed by different QN topology. Simply adding a nonscalable quantum strategy---which can only change the network topology locally---into the QEP scheme is not helpful for making a statistical argument on the percolation threshold, and hence the generality of the quantum advantage on arbitrary network topology is yet to be understood.

%\renewcommand*{\thefootnote}{\fnsymbol{footnote}}
%\begingroup
%\squeezetable
\begin{table*}[t]
	\centering
	\caption{\label{table_thresholds}
		ConPT predicts the lowest threshold compared to those obtained from known  classical-percolation-theory-based schemes. All thresholds are given in $\theta$ under a change of variables 
		$p \equiv 2\sin^2\theta$.
		\protect\footnotetext[1]{\mbox{$P_\text{swap}(k)=2x-x^2$, where $x(k)$ is the solution of $2x+x^k(xk-x-k-1)-(1-x)/(k-1)=0$ by the $q$-swapping strategy~\cite{QEP-q-swap_cc09}.}}
		%		\protect\footnotetext[2]{\mbox{QEP produces the same thresholds as of CEP but shows advantages when the lattice symmetry is broken~\cite{QEP-detail_pcalw08}.}}
		\protect\footnotetext[2]{\mbox{$P_\text{GHZ}(k)=x$ is the solution of $1-(1-x)\sum_{i=0}^{\lfloor k/2-1\rfloor}{\binom{2i}{i} 4^{-i}(2x-x^2)^i}-1/(k-1)=0$ where $\lfloor \cdot\rfloor$ is the floor function~\cite{QEP-GHZ_pclla10}.}}
		\hfill\hfill}
	\medskip
	\vspace{-3mm}
	\begin{tabular}{p{3.2cm}|p{5.0cm} p{2.3cm} p{2.9cm} p{2.4cm}}
		\hline
		[unit: $\left(\pi/4\right)^{-1}\theta$] & Bethe lattice (degree $k$)& Square  & Honeycomb  & Triangular \\
		\hline
		%$p_{\text{th}}\approx0.505$, $p_{\text{th}}\approx0.657$, $p_{\text{th}}\approx0.345$
		CEP~\cite{QEP_acl07}
		%		& $0.526$ &$0.610$ & $0.428$ \\
		&$(4/\pi)\sin^{-1}[1/{\sqrt{2\left(k-1\right)}}]$ & $0.670$ &$0.777$ & $0.545$ \\
		%\hline
		QEP~\cite{QEP_acl07,QEP-q-swap_cc09,QEP-detail_pcalw08} 
		%		& -- & $0.598$ & -- \\
		&$(4/\pi)\sin^{-1}\sqrt{P_\text{swap}(k)/2}$\footnotemark[1] & $0.670$ & $0.761$ & $0.545$\\
		%\hline
		QEP-GHZ~\cite{QEP-GHZ_pclla10} 
		%		& $0.459$ & $0.585$ & $0.378$ \\
		&$(4/\pi)\sin^{-1}\sqrt{P_\text{GHZ}(k)/2}$\footnotemark[2] & $0.584$ & $0.745$ & $0.481$ \\
		%\hline
		ConPT (Fig.~\ref{fig_critical-curve})
		%		& $0.327$ & $0.399$ & $0.250$ \\
		&$(2/\pi)\sin^{-1}(1/{\sqrt{k-1}})$ 
		%	& $0.416$ & $0.508$ & $0.318$
		& $0.42(8)$ & $0.51(8)$ & $0.32(8)$ \\
		\hline
	\end{tabular}
\end{table*}
%\endgroup

In response to the question, 
here we introduce an {alternative mapping} called concurrence percolation theory (ConPT) which explicitly {\emph{relaxes}} the necessity of establishing a path of singlets. 
We directly generalize percolation theory in terms of path connectivity and apply it to \emph{concurrence}~\cite{concurrence_hw97} (a key measure of bipartite entanglement defined as $c:=\sin2\theta$ for a pure state), 
rather than singlet conversion probability like in the traditional CEP/QEP scheme~\cite{QEP_acl07}. The existence of ConPT itself, as we will see, implies that entanglement transmission can also be established when the two infinitely distant nodes are connected by paths of only imperfectly entangled states---as long as there are enough paths. Interestingly, we find that the threshold predicted by ConPT is the lowest threshold compared to earlier known schemes (Table~\ref{table_thresholds}).  
Our results help extending our knowledge of quantum advantage as well as discovering potentially new criticality in entanglement statistics.

\textit{Percolation as a theory of connectivity.---}{Recent results~\cite{QEP-complex-netw_plac10} hint that the cluster size may be an ill-defined order parameter of a genuine statistical theory of entanglement transmission.
Indeed, percolation theory was initially about path connectivity before being reformulated into clusters due to mathematical convenience.
Thus, instead of clusters, here we make direct use of the classical \textit{``sponge-crossing'' probability} $P_{\text{SC}}$---the probability that there is an open path connecting two far-apart boundaries, which was used in the early studies of bond percolation on 
2D (and higher-dimensional) lattices~\cite{cross-probab-square_k80,*cross-probab-triangle_w81}. 
$P_{\text{SC}}$ can be calculated by \emph{connectivity rules} using the link weights $p$ ($0\le p\le1$)---which are simply determined numbers before \emph{a posteriori} explained as occupation probabilities---along all paths that connect the two boundaries.
In the thermodynamic limit, $n\rightarrow\infty$, we expect that $P_{\text{SC}}$ should approach either $0$ or $1$, respectively, in the sub- or supercritical regimes, separated by the percolation threshold $p_\text{th}$~\cite{cross-probab-square_k80,*cross-probab-triangle_w81}.
}

For a series-parallel network~\cite{series-parallel-netw_d65}, by definition, only two connectivity rules, namely, series and parallel rules, are sufficient for calculating $P_{\text{SC}}$. Surprisingly, the series and parallel rules for classical percolation are simple but both are extensible and commutable (Table~\ref{table_rules}), similar to calculating the net electrical resistance in a resistance network. When ``loops'' exist (for example, in a bridge circuit~\cite{series-parallel-netw_d65}), also required are higher-order connectivity rules which are complicated (but closed form owing to the additivity of probability measure). Additionally, these rules can be well approximated by \emph{only} series and parallel rules via a useful technique known as the \emph{star-mesh (SM) transform}~\cite{star-mesh_v70} (Table~\ref{table_rules}), which is similar to a local renormalization group process (see Supplemental Material~\footnote{See Supplemental Material below for a description of star-mesh transform and other results.}). {This technique was used in, e.g.,~the Frank-Lobb algorithm~\cite{frank-lobb_fl88}, for solving classical percolation problems.}

We expect that ConPT can be built similarly, yet not on probability but on concurrence. 
We denote by $C_{\text{SC}}$ the sponge-crossing concurrence and $c_{\text{th}}$ the corresponding threshold on the concurrence $c$ of each link in $\mathcal{G}_{\theta}\left(n\right)$.
$C_{\text{SC}}$ in the sub- or supercritical regimes should also approach either $0$ or $1$ in the thermodynamic limit.
We proceed by examining possible connectivity rules in QN for transmission of concurrence that are allowed by LOCC in an optimal manner. In general, a full probabilistic argument should be built since LOCC involves selective measurements~\cite{q-dissipative-syst} of quantum states and results in probabilistic outcomes~\cite{QEP-detail_pcalw08}. However, there is a subset of LOCC which is considered ``deterministic'' as it only yields one possible outcome in terms of pure states, up to unitary equivalence. The deterministic LOCC is what we need for building connectivity rules so as to keep ConPT a determined theory of connectivity and avoid mixing concurrence with probability measures. Fortunately, we find that ConPT also admits similarly simple but general series and parallel rules~(Table~\ref{table_rules}), the realizability of which by LOCC is discussed below.

\begin{table}[h]
	\centering
	\caption{\label{table_rules}Connectivity rules.}
		\begin{tabular}{p{1.6cm}| p{3.0cm}| p{3.4cm}}
		\hline
		& Classical & ConPT \\
		\hline
		Series rule & $p=p_1 p_2\cdots$ & $c=c_1 c_2\cdots$ \\
		\hline
		Parallel rule & $1-p=$\newline
		$\left(1-p_1\right)\left(1-p_2\right)\cdots$ &  $\frac{1+\sqrt{1-c^2}}{2}=\max\{\frac{1}{2},$\newline
		$\frac{1+\sqrt{1-c_1^2}}{2} \frac{1+\sqrt{1-c_2^2}}{2}\cdots\}$ \\
		\hline

Higher-order~rules & 
\multicolumn{2}{p{6.7cm}}{
%	\vspace{-0.7mm}
	 Can be approximated by the \emph{SM transform} by the following two-step argument:
}
\\
\cline{1-1}
\end{tabular}
\begin{tabular}{p{5.8cm}p{2.2cm}}
\vspace{-1.0mm}
1. The SM transform can reduce an $n$-graph to an $(n-1)$-graph (right panel) and is solvable by applying the series and parallel rules recursively through a group of $n(n-1)/2$ coupled equations (see Supplemental Material for details).
\newline
2. Applying the transform consecutively on a network can reduce nodes one by one---and thus reduce any topology to two nodes, yielding the final (approximate) connectivity between them (bottom panel, i.~$\to$~viii.).
&
\begin{center}
	\vspace{-5mm}
	{\includegraphics[width=2.4cm]{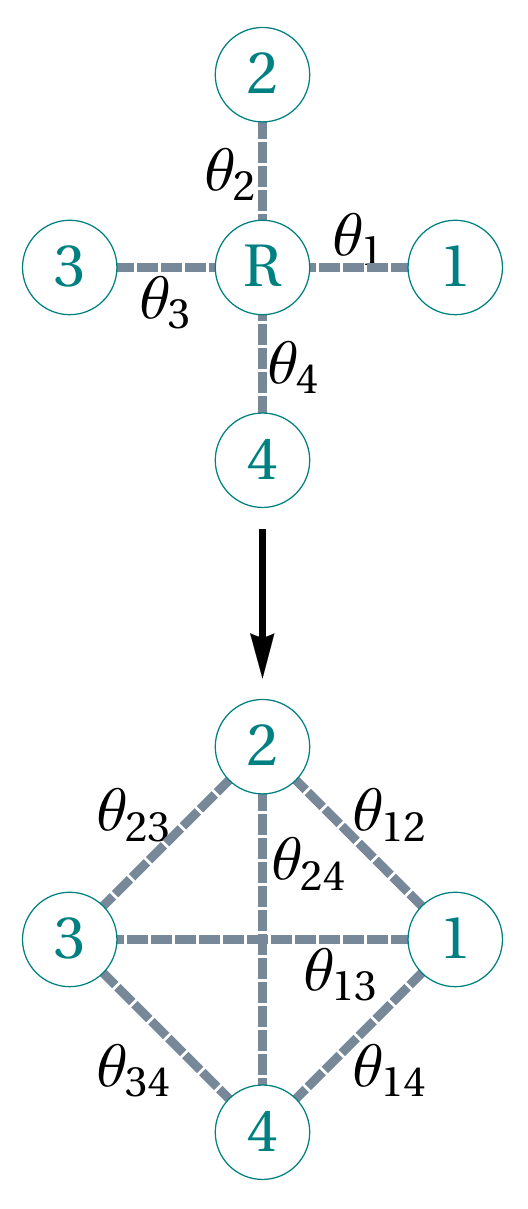}}
\end{center}
\\	
\multicolumn{2}{p{8.0cm}}{
	\begin{center}	
		\vspace{-7mm}	
		{\includegraphics[width=8.3cm]{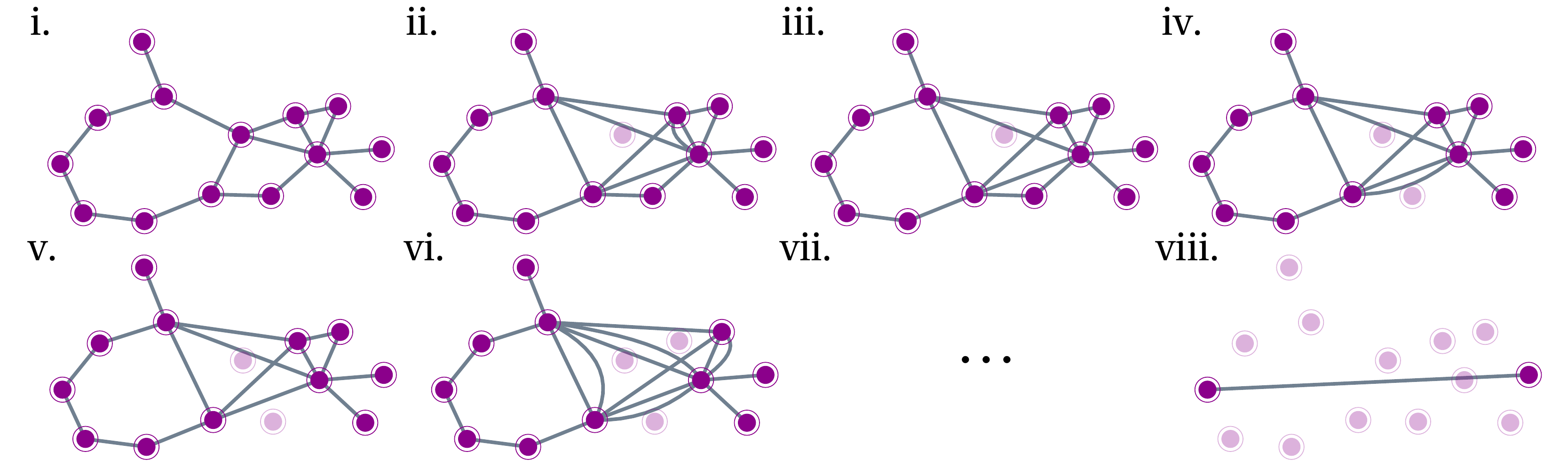}}
		\vspace{-7mm}
	\end{center}
}  \\		

		\hline
	\end{tabular}
\end{table}

\textit{Series and parallel rules as LOCC.---}(i)~Series rule. When two links of concurrence $c_\text{AR}$, $c_\text{RB}$ are connected in series between three nodes, Alice-Relay-Bob (A-R-B), ``swapping'' on R projects out four probabilistic outcomes between A and B~\cite{entangle-swap_zzhe93,*QEP-series-rule_bvk99}. The final average concurrence is $C=\sum^{4}_{k=1}{\omega_k C_k}$ where $\omega_{k}$ is the probability of producing a pure state of concurrence $C_k$. $\sum_k{\omega_{k}}=1$ is understood.
When a particular Bell basis (the \emph{XZ}~basis~\cite{QEP-detail_pcalw08}) is chosen for projection, not only is $C$ optimal but also all $C_k$ are identical to the product of concurrences of the two links, $C_k=c_\text{AR}c_\text{RB}$, hence admitting deterministic LOCC. (ii) Parallel rule. For two parallel links between A and B, the product state
$\left|\psi_\text{AB}(\theta_1)\right\rangle\otimes\left|\psi_\text{AB}(\theta_2)\right\rangle
=(\cos\theta_1\left|00\right\rangle+\sin\theta_1\left|11\right\rangle)(\cos\theta_2\left|00\right\rangle+\sin\theta_2\left|11\right\rangle)
$
belonging to $\mathcal{H}_1\otimes\mathcal{H}_2$ is a ``two-ququart'' state. 
By Nielsen's theorem~\cite{nielsen_n99}, the maximally entangled two-qubit pure state obtainable by LOCC is $\cos\theta_\text{tot}\left|00\right\rangle+\sin\theta_\text{tot}\left|11\right\rangle$, where $\cos\theta_\text{tot}=\cos\theta_1\cos\theta_2$ is equal to the largest Schmidt coefficient, provided that $\cos\theta_1\cos\theta_2>1/\sqrt{2}$. When $\cos\theta_1\cos\theta_2\le1/\sqrt{2}$, a singlet $\cos\theta_\text{tot}=1/\sqrt{2}$ can always be obtained. Again, not only is the LOCC deterministic but it actually optimizes the obtainable average concurrence $C=\sum_{k}{\omega_k C_k}$ %=\sin2\theta_\text{tot}$ 
as well, a result of concurrence being an entanglement monotone~\cite{QEP-parallel-rule_v99,*monotone_v00}.

A particular realization of these LOCC on some series-parallel hierarchical lattices~\cite{hierarchy-netw_rha07,*hierarchy-netw_ra07} has been given in Ref.~\cite{QEP-detail_pcalw08}, where the series rule is called a worst-case entanglement (WCE) strategy, since it maximizes the WCE established in a 1D chain. 
Here, we argue that the parallel rule is also a WCE strategy for parallel links, because it not only maximizes the average concurrence but also guarantees that the worst case is equal to the average.

On general networks, the higher-order connectivity rules produced by the SM transform may not be realizable by LOCC. They are only approximations of the true LOCC-allowing rules. Generalizing a quantum channel by including multiple entanglement links may help us understand and even determine the true rules---a difficult task to be handled by multipartite strategies~\cite{QEP-GHZ_pclla10} and QN routing~\cite{q-netw-route_p19,*q-netw-route_pkttjbeg19}.

\begin{figure}[t!]
	\centering
	\begin{minipage}[b]{24mm}	
		\centering
		{\includegraphics[height=17mm]{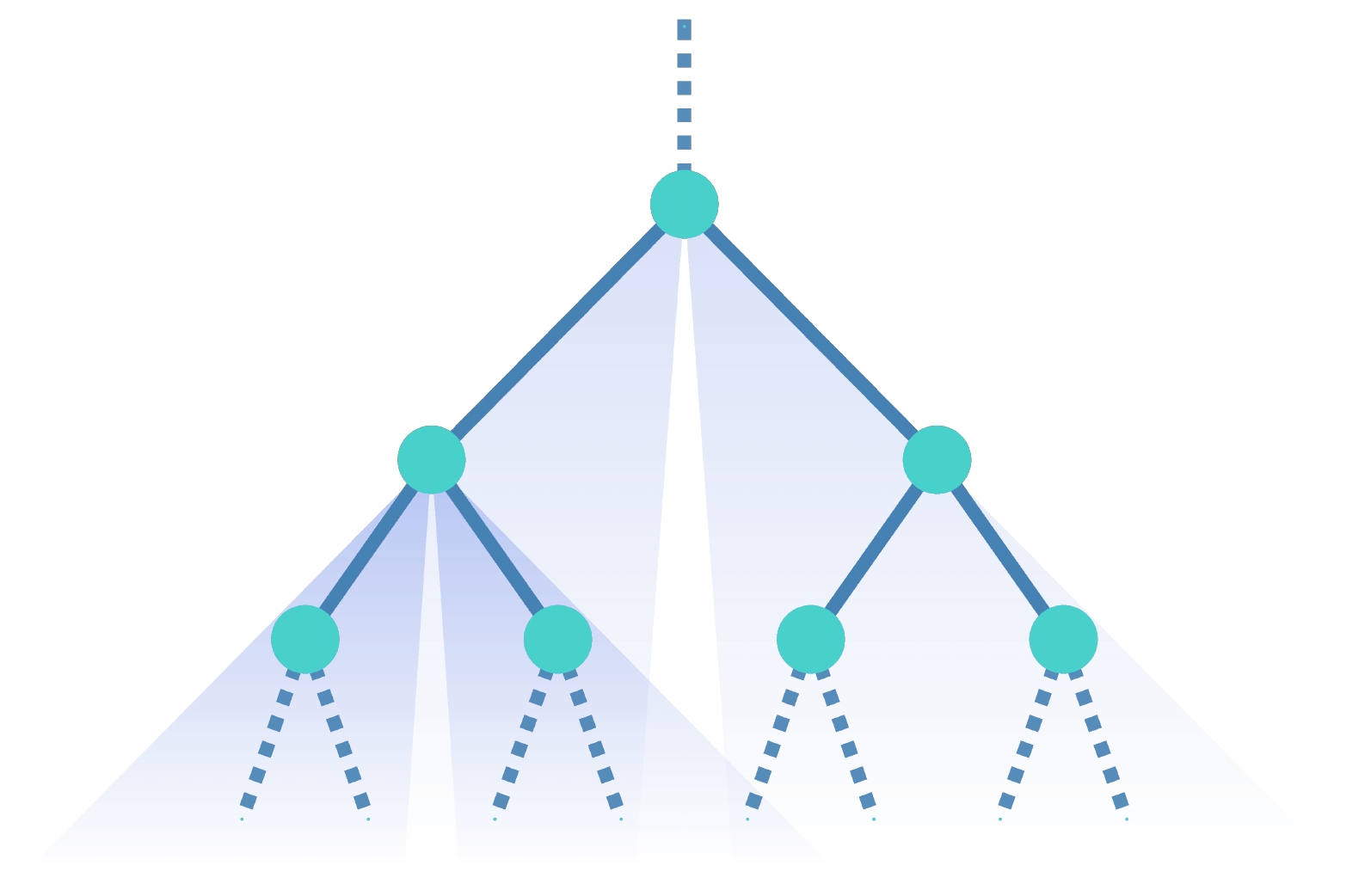}
			\vspace{-6mm}\subcaption{\label{fig_lattice_bethe}}}
	\end{minipage}
	\begin{minipage}[b]{19mm}
		{\includegraphics[height=17mm]{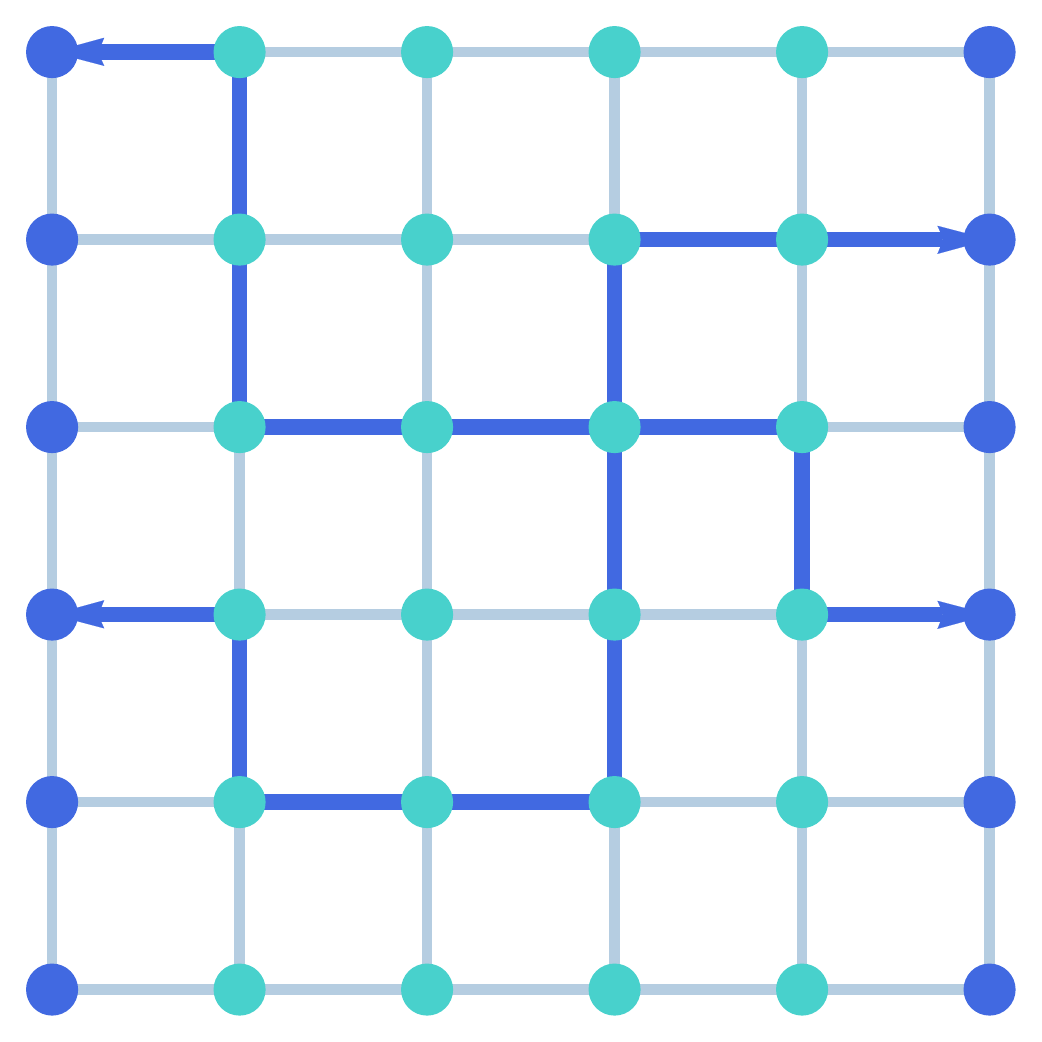}
			\vspace{-2mm}\subcaption{\label{fig_lattice_square}}}
	\end{minipage}
	\begin{minipage}[b]{19mm}
		{\includegraphics[height=17mm]{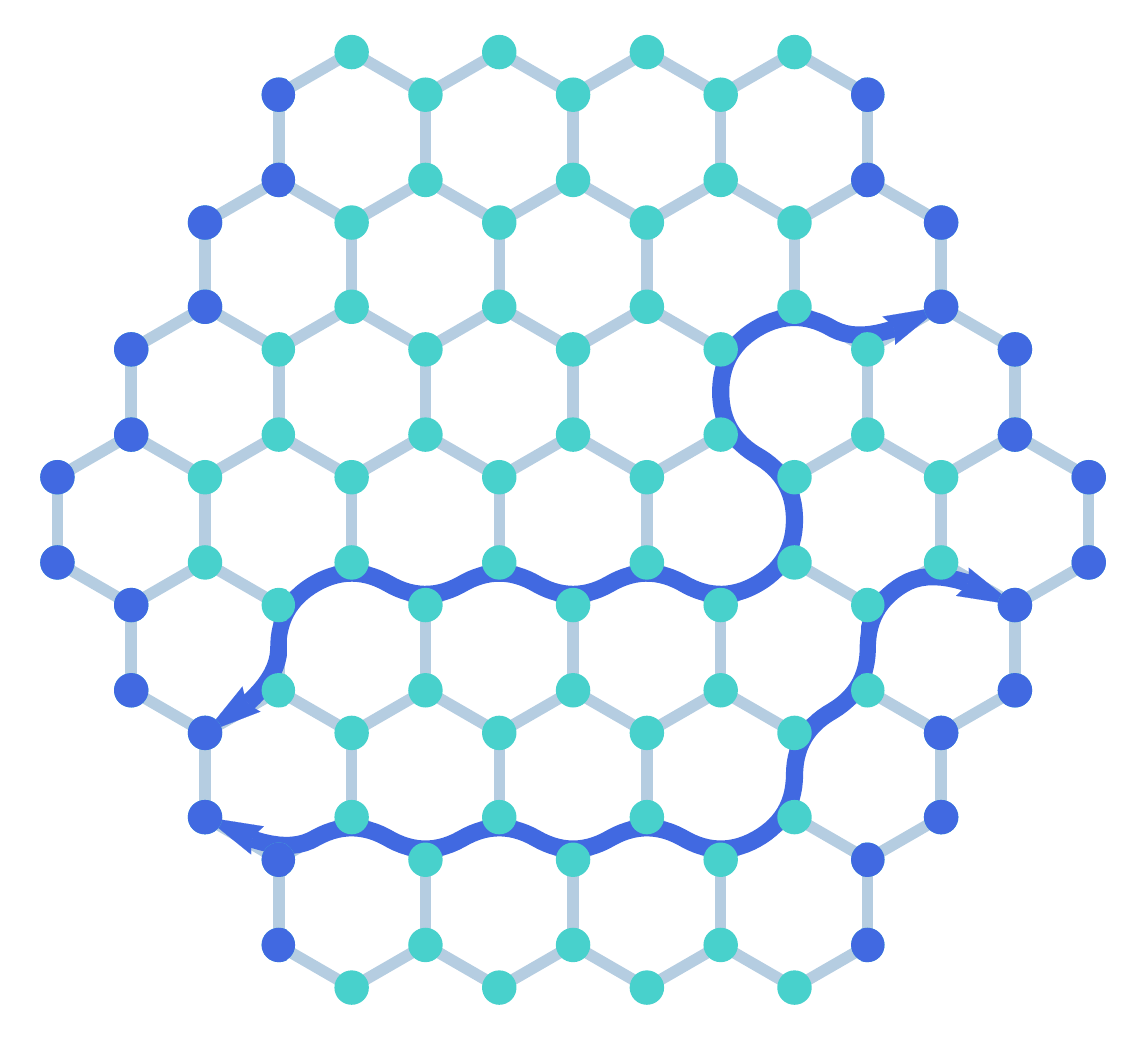}
			\vspace{-6mm}\subcaption{\label{fig_lattice_hex}}}
	\end{minipage}
	\begin{minipage}[b]{19mm}
		{\includegraphics[height=17mm]{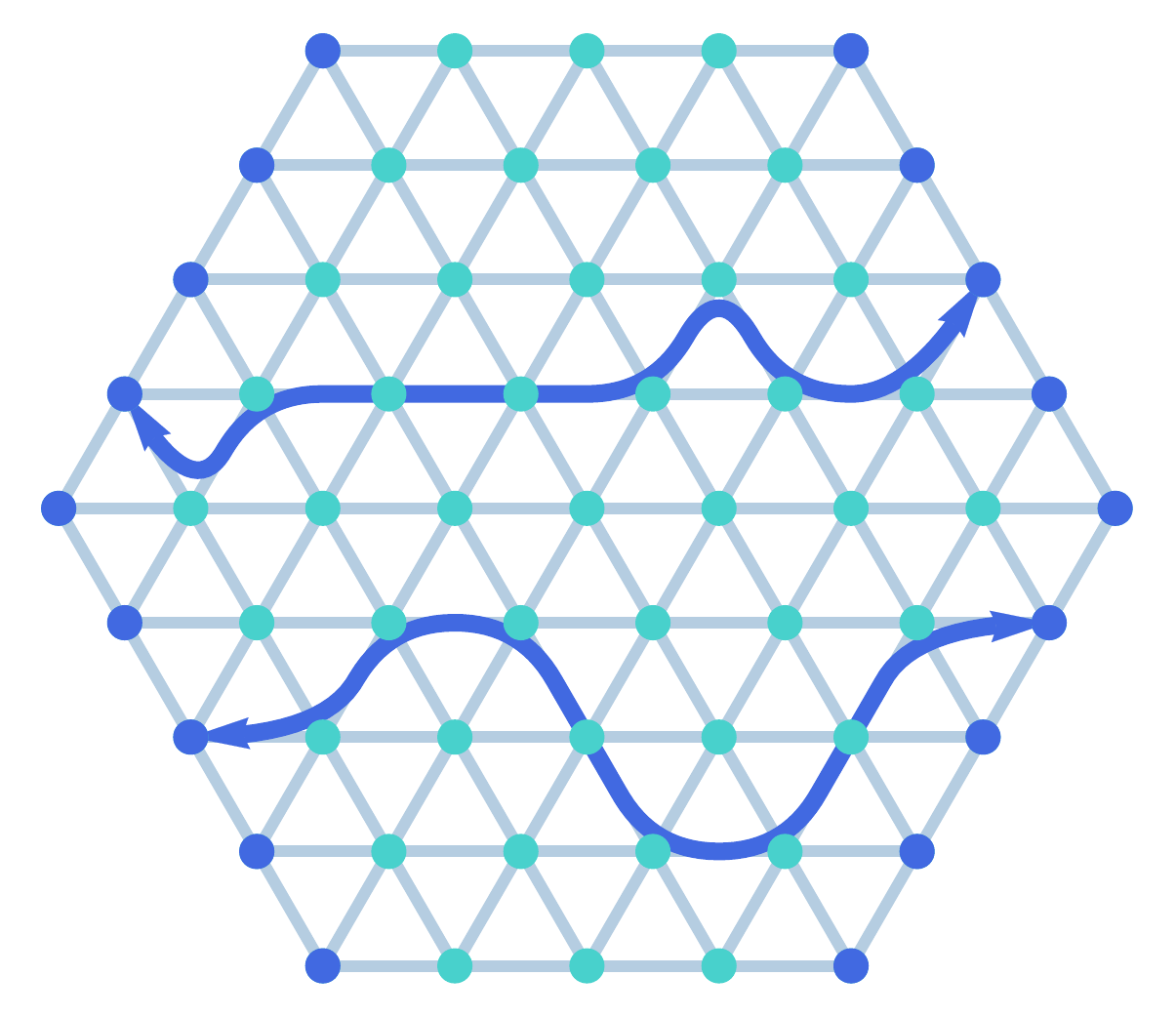}
			\vspace{-6mm}\subcaption{\label{fig_lattice_tri}}}
	\end{minipage}
	
	\begin{minipage}[b]{42mm}
		\centering
		{\includegraphics[width=42mm]{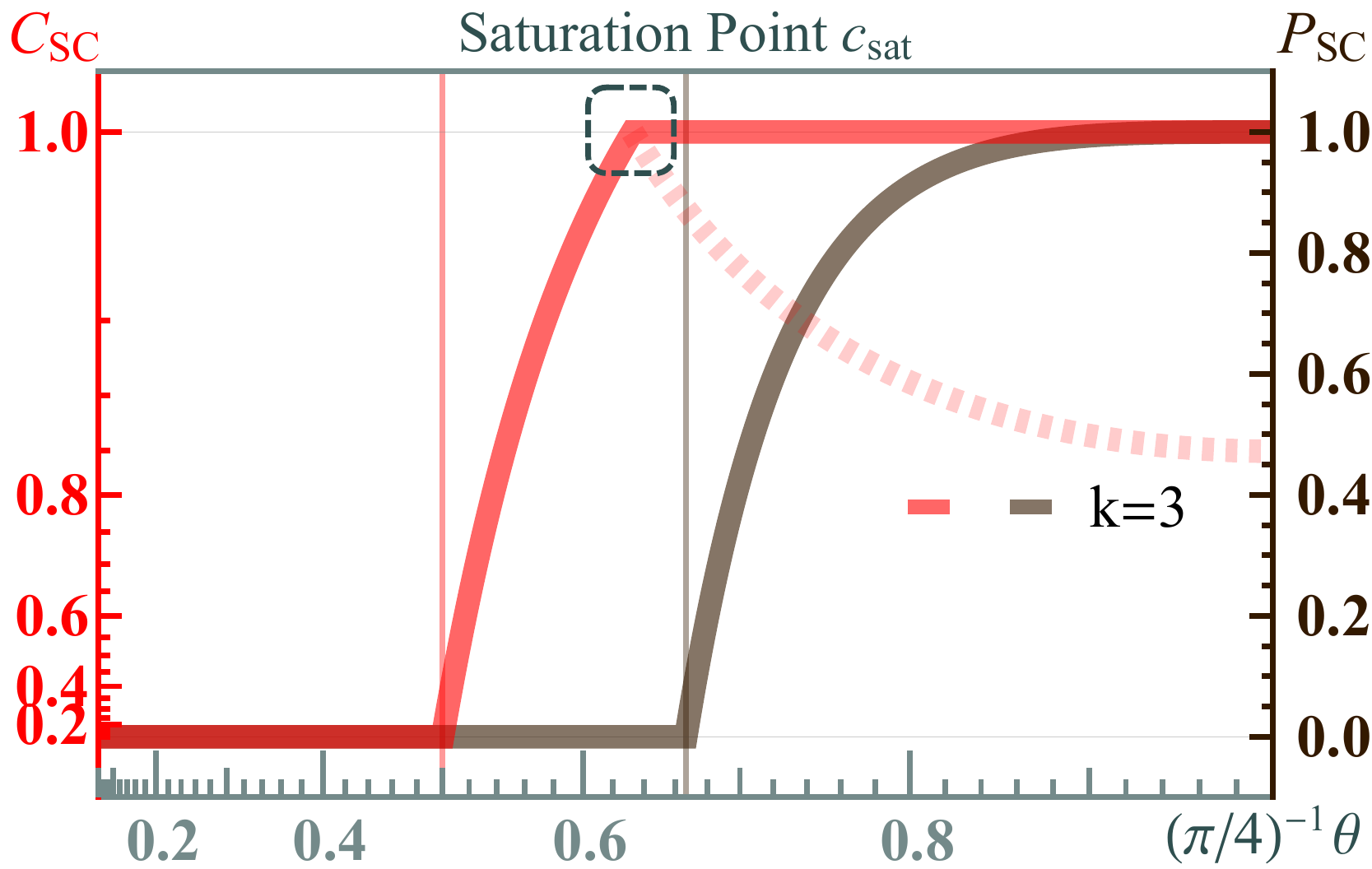}
			\vspace{-7.5mm}
			\subcaption{\label{fig_critical-curve_bethe}}}
		%				\addtocounter{subfigure}{-4}
		\vspace{-3mm}
		{\includegraphics[width=42mm]{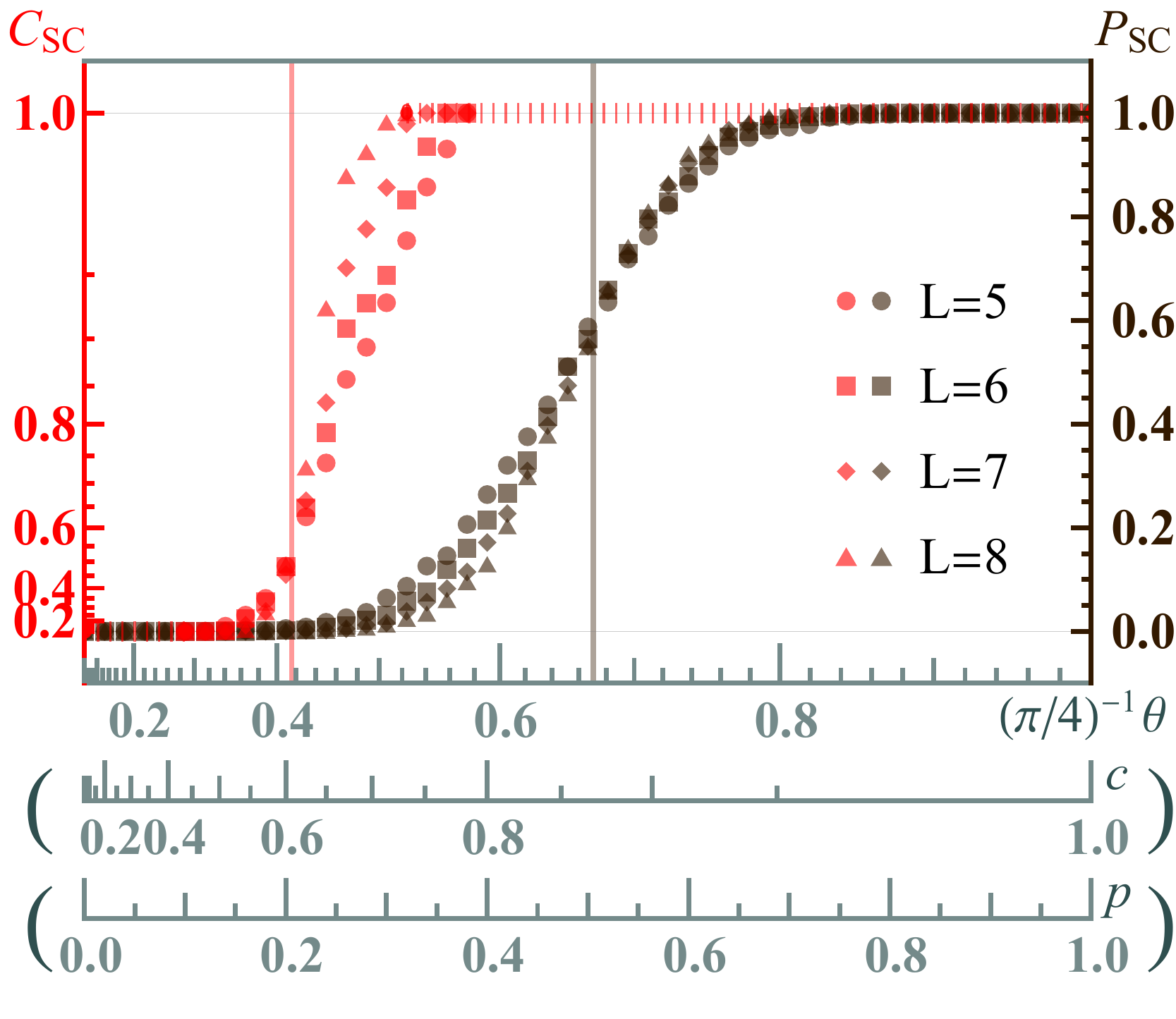}
			\vspace{-8.5mm}
			\subcaption{\label{fig_critical-curve_a}}}
		%				\addtocounter{subfigure}{1}
	\end{minipage}
	\begin{minipage}[b]{42mm}
		\centering
		{\includegraphics[width=42mm]{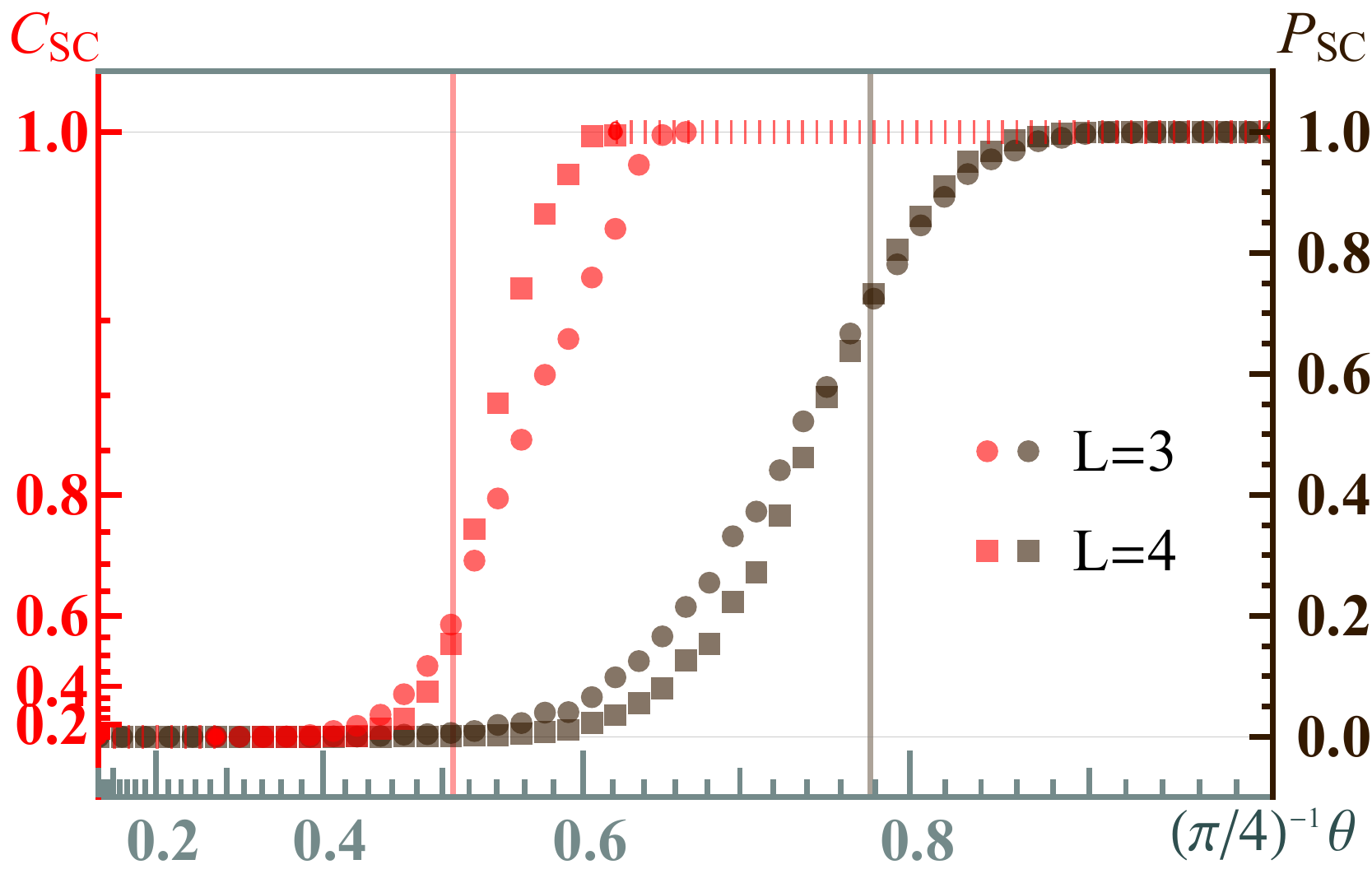}
			\vspace{-7.5mm}
			\subcaption{\label{fig_critical-curve_b}}}	
		%				\addtocounter{subfigure}{1}
		\vspace{-3mm}
		{\includegraphics[width=42mm]{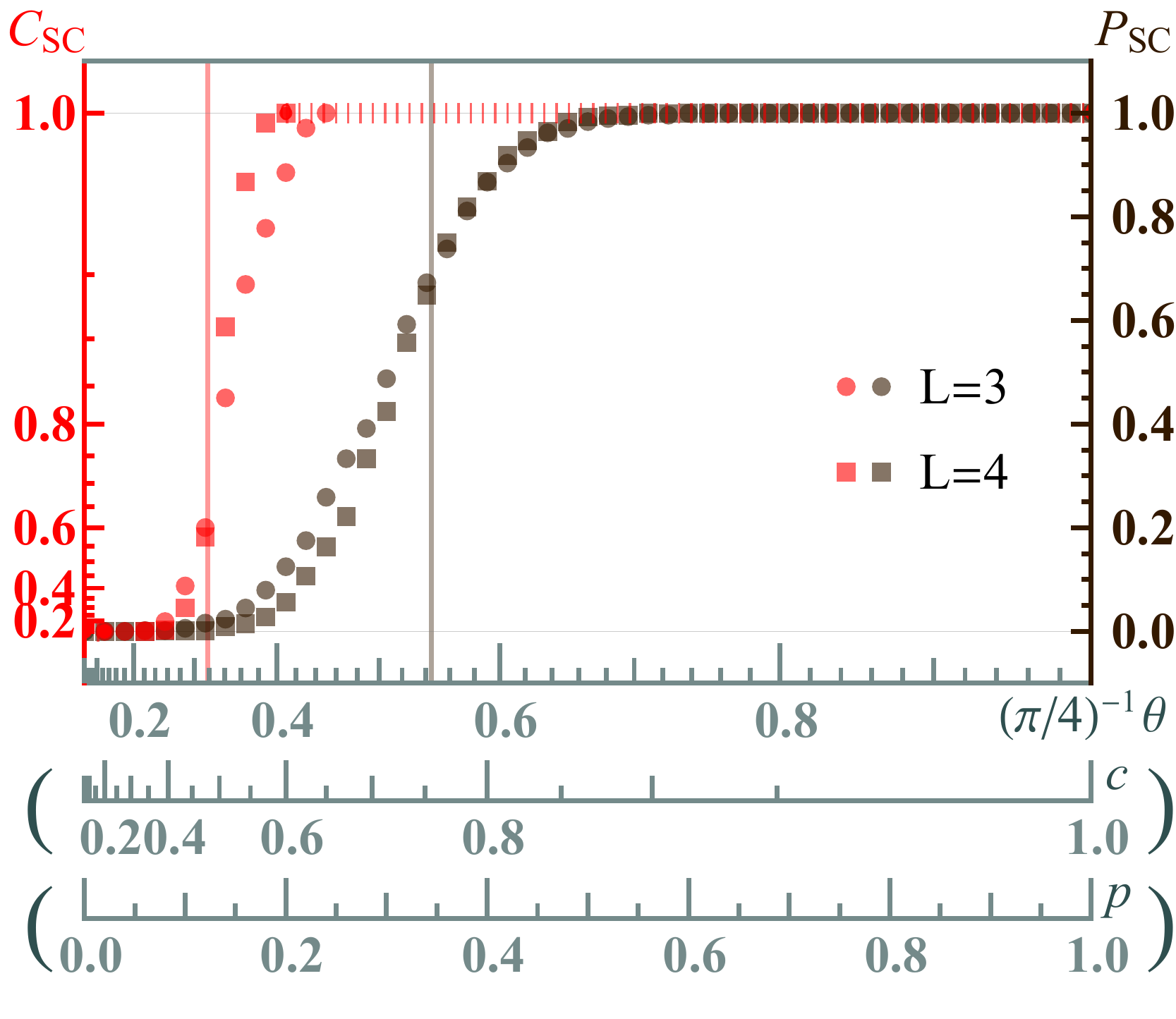}
			\vspace{-8.5mm}
			\subcaption{\label{fig_critical-curve_c}}}
	\end{minipage}
	\caption{\label{fig_critical-curve}Comparison between classical percolation theory and ConPT.
		\subref{fig_lattice_bethe}~Bethe lattice (e.g.,~$k=3$). \subref{fig_lattice_square}~Square lattice (e.g.,~\mbox{$L=5$}). \subref{fig_lattice_hex}~Honeycomb lattice (e.g.,~\mbox{$L=4$}). \subref{fig_lattice_tri}~Triangular lattice (e.g.,~\mbox{$L=4$}). 
		\subref{fig_critical-curve_bethe}~For the Bethe lattice, $P_{\text{SC}}=1-(1-p)^3/p^3$ yields $p_\text{th}=1/2$ for $k=3$; $C_{\text{SC}}=\sin\{2\cos^{-1}[({\sqrt{1/4+c^{-2}} }-1/2)^{3/2}]\}$ yields not only $c_\text{th}=1/\sqrt{2}$ but also $c_\text{sat}$ above which the analytical solution is unphysical (red dashed), making $C_\text{SC}=1$ when $c\ge c_\text{sat}$.
		\subref{fig_critical-curve_a}--\subref{fig_critical-curve_c}~For the corresponding 2D lattice types \subref{fig_lattice_square}--\subref{fig_lattice_tri}, SM transform approximations produce $C_{\text{SC}}$ (red) with respect to $c\equiv \sin2\theta$, compared with $P_{\text{SC}}$ (brown) with respect to $p\equiv 2\sin^2\theta$ produced by standard Monte Carlo simulations. 
		$c_{\text{th}}$ (red vertical) and $p_{\text{th}}$ (brown vertical) are determined by their finite-size crossing points.	\hfill\hfill}
\end{figure}

\textit{Percolation thresholds.---}The Bethe lattice is a typical series-parallel network where each node has the same degree $k$  [Fig.~\ref{fig_lattice_bethe}]. $P_{\text{SC}}$ and $C_{\text{SC}}$ are defined as 
between the root and the entire boundary and can be solved exactly. 
Using an exact renormalization technique on the series and parallel rules~(see Supplemental Material), we first recover the classical threshold $p_{\text{th}}=1/(k-1)$; whereas in ConPT we find $c_{\text{th}}=1/\sqrt{k-1}$, and thus the ConPT threshold is always smaller, i.e.,~$1-\sqrt{1-c_\text{th}^2}\le p_\text{th}$. 
Interestingly, the percolation curve of $C_{\text{SC}}$ [Fig.~\ref{fig_critical-curve_bethe}] exhibits not only a percolation threshold $c_\text{th}$ but also a saturation point $c_{\text{sat}}$ which can be solved exactly too, $c_{\text{sat}}=\sqrt{(1/2)^{1/k}-(1/4)^{1/k}}/\sqrt{(1/2)^{(k-1)/k}-(1/4)^{(k-1)/k}}$, an anomaly of the ConPT parallel rule (Table~\ref{table_rules}) being not a smooth function. 
The existence of a saturation point reflects a stunning quantum advantage in Bethe lattices: with certainty one can establish a singlet that connects any node to the boundary, as long as the entanglement in each link exceeds the saturation point. This advantage cannot be revealed from any scheme based on classical percolation theory where a singlet can only be established with certainty if each link is also perfectly entangled.

{If we replace $k$ by $fk+(1-f)$ in $p_\text{th}$ and $c_\text{th}$ ($0<f\le1$), then $p_\text{th}$ and $c_\text{th}$ will denote the thresholds not for the original Bethe lattice but for a diluted one where $1-f$ fraction of links are randomly removed (see Supplemental Material). A less-than-one $f$ can be understood as an imperfection of LOCC, and the dependence of $p_\text{th}$ and $c_\text{th}$ on $f$ thus determines the robustness of entanglement transmission under random imperfections. When $f<1/(k-1)$, both $p_\text{th}$ and $c_\text{th}$ become unphysical because of the breakdown of the Bethe lattice structure.}

Finally, Figs.~\ref{fig_lattice_square}--\ref{fig_lattice_tri} show 2D lattices with left and right boundaries (blue dots) and possible paths connecting them (arrow lines), for which the SM transform must be used to determine the higher-order connectivity. Shown correspondingly in Figs.~\ref{fig_critical-curve_a}--\ref{fig_critical-curve_c} are how the sponge-crossing quantities change as a function of $p$ and $c$.
We find, again, that the thresholds predicted by ConPT are always smaller. 
Indeed, this result can be understood in an exact manner by directly comparing the series and parallel rules in Table~\ref{table_rules}~{\footnote{{Bear in mind that $1-c_i^2\equiv(1-p_i)^2$. Comparing the series rules yields
$c^2=\prod_i c_i^2=\prod_i p_i\left(2-p_i\right)\ge\left(\prod_i p_i\right) \left(2-\prod_i p_i\right)=p\left(2-p\right)$ which is proved by the subadditivity of $f(x)=\ln(2-e^{-x})$ for $x\ge0$. Comparing the parallel rules yields
$\frac{1}{2}+\frac{1}{2}\sqrt{1-c^2}=\prod_i ({\frac{1}{2}+\frac{1}{2}\sqrt{1-c_i^2}})=\prod_i\left(1-\frac{p_i}{2}\right)\le\frac{1}{2}+\frac{1}{2}\prod_i\left(1-p_i\right)=1-\frac{p}{2}$ which is proved by the subadditivity of $f(x)=-\ln(1/2+e^{-x}/2)$ for $x\ge0$. Both inequalities further yield $1-\sqrt{1-c^2}\ge p$, showing the general quantum advantage independent of QN topology in both the series and parallel rules of ConPT.} }}.

\begin{figure}[t!]
\centering
\begin{minipage}[b]{40mm}
	\centering
	{\includegraphics[height=29mm]{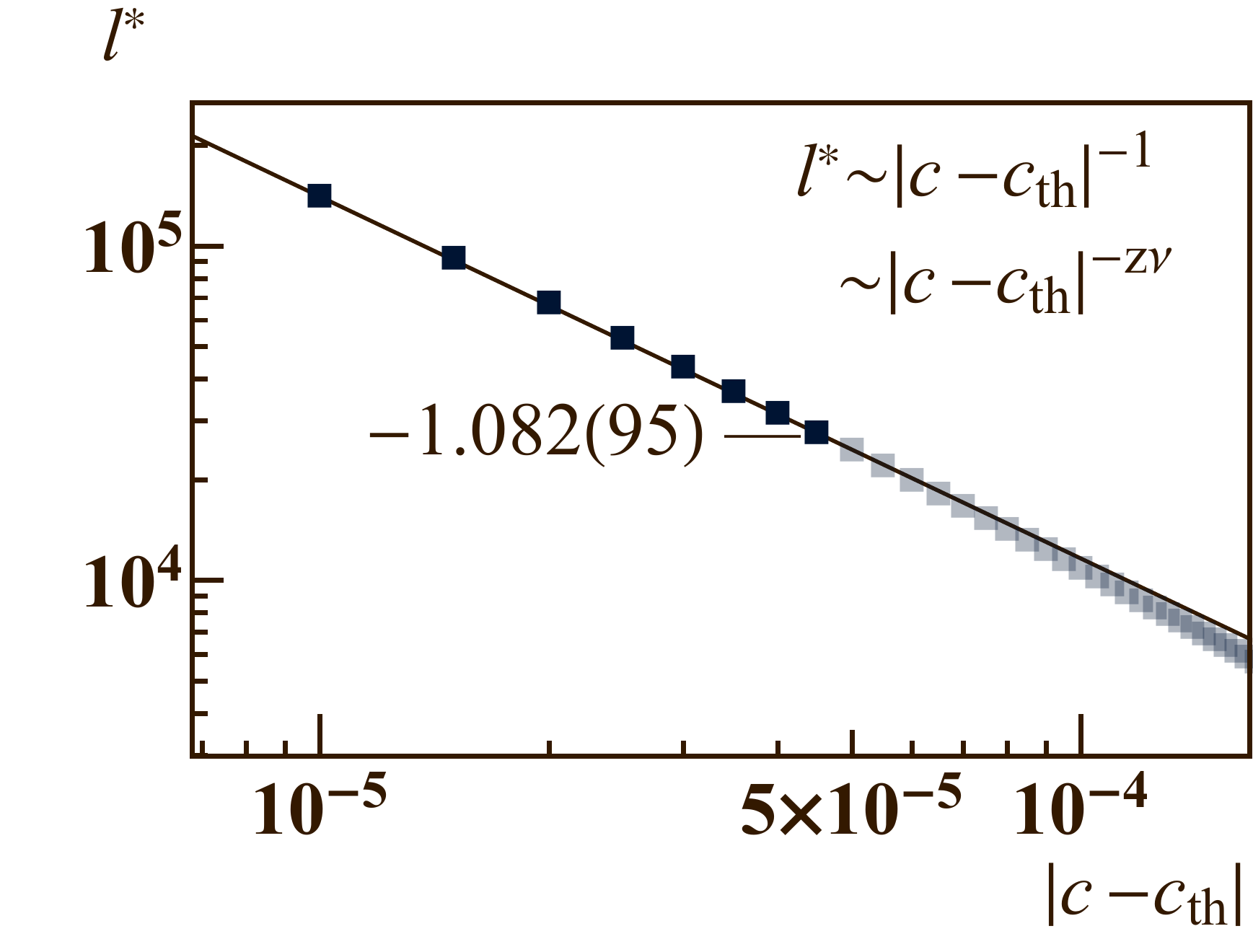}
		\vspace{-7.5mm}
		\subcaption{\label{fig_bethe_scaling_c11}}}
\end{minipage}
\begin{minipage}[b]{45mm}
	\centering
	{\includegraphics[height=29mm]{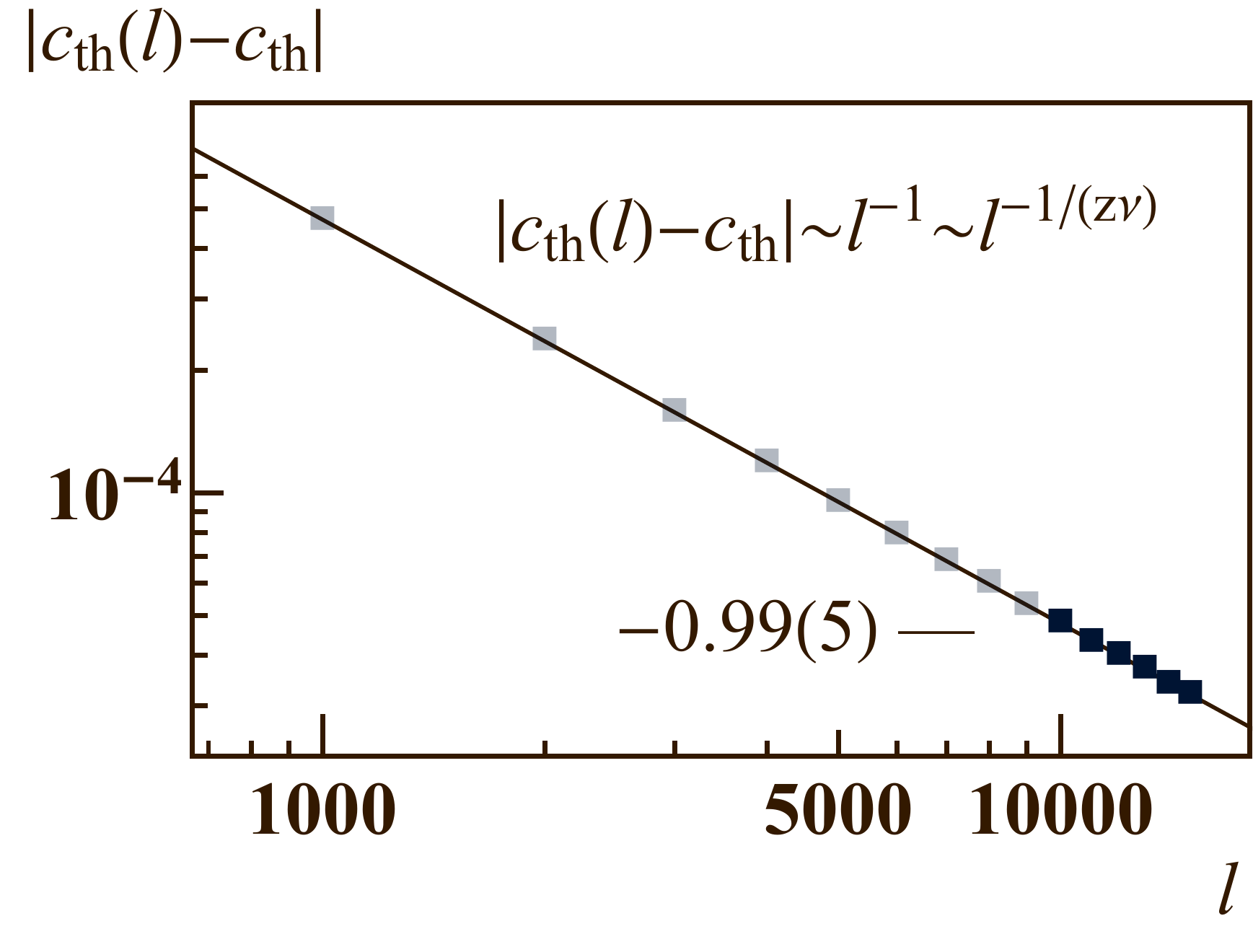}
		\vspace{-7.5mm}
		\subcaption{\label{fig_bethe_scaling_c21}}}
\end{minipage}
\begin{minipage}[t]{84mm}
	\centering
	\vspace{-5mm}
	{\includegraphics[height=29mm]{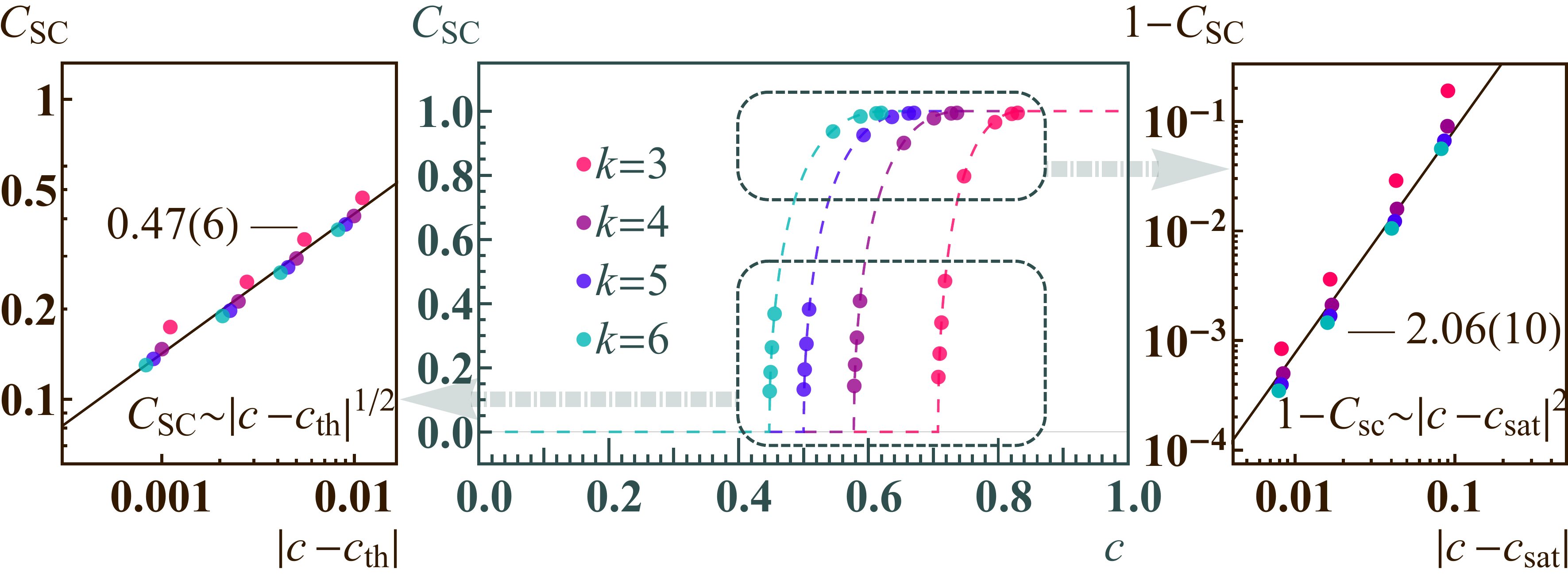}
		\vspace{-5mm}
		\subcaption{\label{fig_bethe_scaling_c3}}}
\end{minipage}
	
	\caption{\label{fig_bethe}
		Universality for the Bethe lattice.
		\subref{fig_bethe_scaling_c11}~Finite-size analysis of ConPT below $c_{\text{th}}=1/\sqrt{2}$ ($k=3$). $C_{\text{SC}}$ follows a power law with an exponential cutoff with respect to the number of layers $l$,
		$C_{\text{SC}}\sim l^{-1/2}\exp(-l/l^*)$, where $l^*$ diverges as a power law when approaching $c_\text{th}$. Numerically $z\nu=1.082(95)$ is obtained by fitting near $|c-c_\text{th}|\sim10^{-5}$ (dark blue squares).
		\subref{fig_bethe_scaling_c21}~Finite-size analysis above $c_\text{th}$ ($k=3$). The finite-size critical threshold $c_\text{th}(l)$ is defined as the turning point of $C_{\text{SC}}$, $c_\text{th}(l)=c|_{\partial^2C_{\text{SC}}/\partial {c}^2=0}$, which deviates from $c_\text{th}$ as a power law with respect to $l$.
		Again, numerically $1/(z\nu)=0.99(5)$ is obtained near $l\sim10^4$ (dark blue squares). \subref{fig_bethe_scaling_c3}~$c_\text{th}$ and $c_\text{sat}$ for general $k$. Two universal power laws of $C_{\text{SC}}$ with respect to $c$ are found by series expansions near $c_{\text{th}}$ and $c_{\text{sat}}$ and confirmed by numerical results (dots) on a finite Bethe lattice of $l=500$.		\hfill\hfill}
\end{figure}

\textit{Critical behavior.---}Percolation theory is associated with universal critical behavior near the percolation threshold. We hypothesize that ConPT as a generalization of bond percolation should also exhibit critical exponents that depend on dimensionality but not on short-range details. 
However, ConPT is not defined by clusters but on paths, thus lacking a suitable clusterlike definition of an order parameter. Hence, we focus solely on the thermal exponent, $\nu$, which characterizes the divergence of correlation length~\cite{diffus-react-fractal-disorder-syst}. $\nu$ is fully and universally determined by the spacial dimension but not the order parameter and therefore may be confirmed conclusively.

The most feasible way to extract $\nu$ is by finite-size analysis. Figure~\ref{fig_bethe} shows how this is done for the Bethe lattice which yields mean-field exponents. Recognizing the number of layers $l$ as the shortest-path distance between the root and the boundary~\cite{diffus-react-fractal-disorder-syst},
we find $z\nu=1$ both below $c_\text{th}$ [Fig.~\ref{fig_bethe_scaling_c11}] and above $c_\text{th}$ [Fig.~\ref{fig_bethe_scaling_c21}] where $z$ is the dynamical exponent.
For infinite-dimensional structures it is reasonable to expect that $l^* \sim \xi^z=\xi^2$ holds as a general random-walk nature~\cite{diffus-react-fractal-disorder-syst} between the characteristic ``time'' $l^*$ and the Euclidean correlation length $\xi\sim\left|c-c_{\text{th}}\right|^{-\nu}$ for not only classical percolation but also ConPT. Thus $\nu=1/2$ is derived.

We may proceed and find other universal power laws, especially, $C_{\text{SC}}\sim|c-c_\text{th}|^{1/2}$ and $1-C_{\text{SC}}\sim|c-c_\text{sat}|^{2}$ near $c_{\text{th}}$ and $c_{\text{sat}}$, respectively, independent of $k$ [Fig.~\ref{fig_bethe_scaling_c3}]. However, in ConPT there is no reason to fix the order parameter to be $C_{\text{SC}}$ and claim that $\beta=1/2$. It is equally possible to let the order parameter be $C_{\text{SC}}$ to some arbitrary $x$th power near the critical threshold, but then we will have $\beta=x/2$ unfixed. 

\begin{figure}[t]
	\centering
	\hspace{-2mm}
	\begin{minipage}[t]{43mm}
		{\includegraphics[width=43mm]{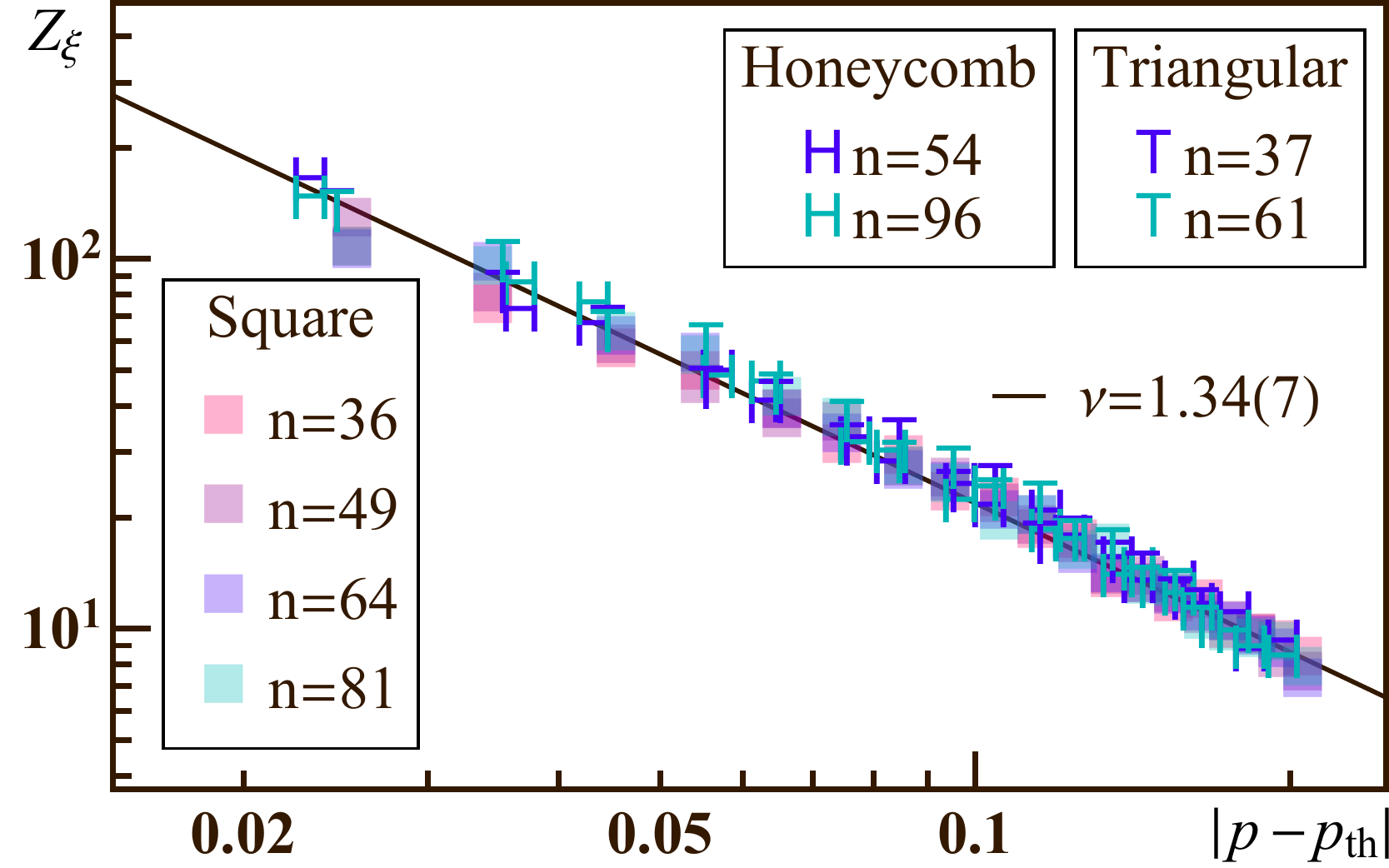}
			\vspace{-6mm}
			\subcaption{\label{fig_2d_scaling_1}}}
	\end{minipage}
	\begin{minipage}[t]{43mm}
		{\includegraphics[width=43mm]{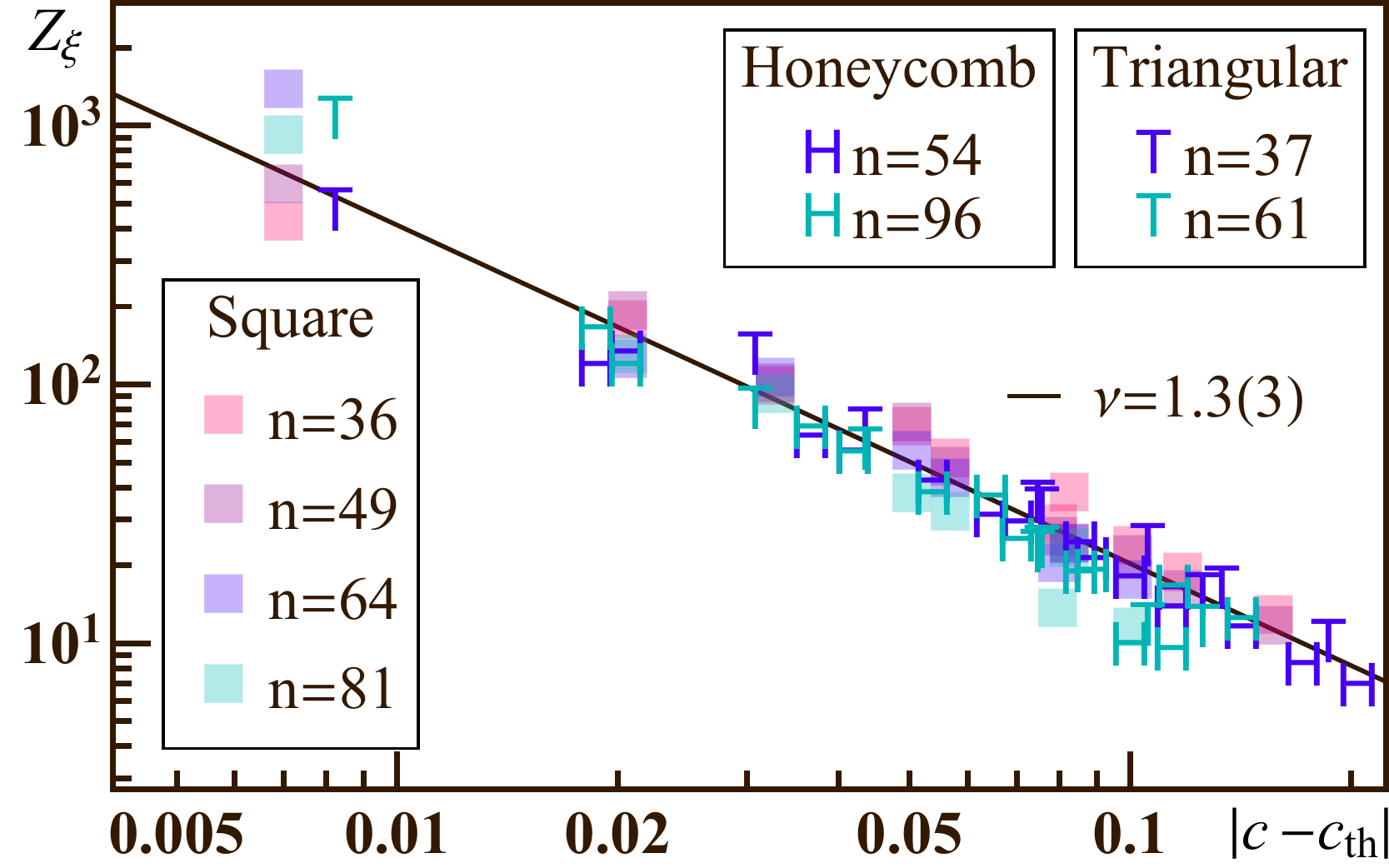}
			\vspace{-6mm}
			\subcaption{\label{fig_2d_scaling_2}}}
	\end{minipage}
	
	\caption{\label{fig_critical-exponent}Universality for 2D lattices. Finite-size analysis is performed in Figs.~\ref{fig_critical-curve_a}--\ref{fig_critical-curve_c}. \subref{fig_2d_scaling_1} For classical percolation $\nu=1.34(7)$ is known and verified. \subref{fig_2d_scaling_2} For ConPT $\nu=1.3(3)$ is found. $Z_{\xi}=\left|p-p_{\text{th}}\right|^{-\nu}$ (or $\left|c-c_{\text{th}}\right|^{-\nu}$).\hfill\hfill}
\end{figure}

Finite-size analysis on 2D lattices is more difficult. As shown in Figs.~\ref{fig_critical-curve_a}--\ref{fig_critical-curve_c}, both $P_{\text{SC}}$ and $C_{\text{SC}}$ seem to gradually converge to a step function as the system size $L\sim\sqrt{n}$ increases, exhibiting an essential finite-size effect (despite very small $L$ because of the heavy computation needed in solving the SM transform). Here, we take advantage of a set of finite-size scaling relations first established by Kesten~\cite{scale-relat_k87,*scale-relat-3d_bcks99,*cross-probab-finite-size-scale_n08} in 2D percolation to explain the universal exponential decay of sponge-crossing probability in either the subcritical regime, given by
$P_{\text{SC}}\left[\mathcal{G}_{\theta}\left(n\right)\right]\sim e^{-L/\xi}$, $p<p_{\text{th}}$,
or the supercritical regime, by
$1-P_{\text{SC}}\left[\mathcal{G}_{\theta}\left(n\right)\right]\sim e^{-L/\xi}$, $p>p_{\text{th}}$.
The scaling of $Z_{\xi}$~\footnote{The universal renormalization coefficient of correlation length $Z_{\xi}$ is defined by $\ln\xi=\ln Z_\xi+\text{nonuniversal terms}$.} in both sub- and supercritical regimes for classical percolation is jointly plotted [Fig.~\ref{fig_2d_scaling_1}]. The thermal exponent hence obtained is close to the known exact and universal value $\nu=4/3$ for different 2D lattices. On the other hand, for ConPT a similar value $\nu=1.3(3)$ is obtained [Fig.~\ref{fig_2d_scaling_2}], also seemingly independent of lattice types. The 2D thermal exponents of both classical percolation and ConPT are thus not very different, hinting that the two might belong to the same universality class. 
This can be eventually tested if a proper definition of order parameter for ConPT is possible so that other critical exponents will be accessible.

\textit{Discussion.---}Our results promote the comprehension of how efficient an entanglement transmission strategy can be designed by LOCC. That being said, it is necessary to expand the theoretical framework to understand mixed states, as any realistic quantum device will unavoidably bring in thermal noise or randomly break down.
Another theoretical interest here is the shift of focus of statistical theory from clusters to paths. This has been considered and explored in problems of classical directed percolation~\cite{direct-percolation_h00} and the corresponding quantum topological order models~\cite{topol-order_dnr89,*topol-order_cn08,*topol-order_cp17,*topol-order_c18,*topol-order_cpt19}. Our results suggest that path connectivity should be more general than clusterlike quantities, as the latter are always limited to probability measures yet the former is not. We hope to understand this better also for higher-dimensional lattices and complex networks in the future.

%\textit{Acknowledgment.---}
We thank I.~Bonamassa, R.~Berkovits, H.~E.~Stanley, and J.~Morrow for their help and useful discussions.
X.~M. would like to thank J.~Ma, T.~Yang, and Y.~Xin for discussion. S.~H. acknowledges the support of Israel Science Foundation, ONR, and BSF-NSF Grant No.~2019740. X.~M. and S.~H. are supported by DTRA Grant HDTRA-1-19-1-0016. J.~G. acknowledges the support of National Science Foundation under Grant No.~2047488, and the Rensselaer-IBM AI Research Collaboration.

\bibliography{ConPT}

\onecolumngrid

\beginsupplement

\clearpage

\tableofcontents
\newpage

\section{The Star-Mesh Transform and the Approximation Procedure}

\begin{figure}[t!]	
	\centering
	\begin{minipage}[b]{8.8cm}
		\centering
		
		\includegraphics[width=8.23cm]{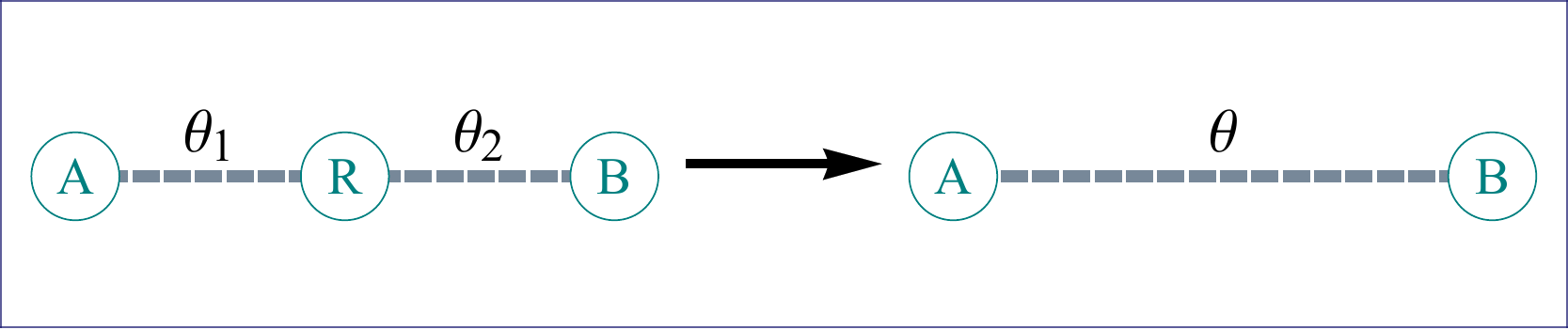}
		\subcaption{\label{fig_rules_a}}
		\vspace{5.7mm}
		
		\includegraphics[width=8.23cm]{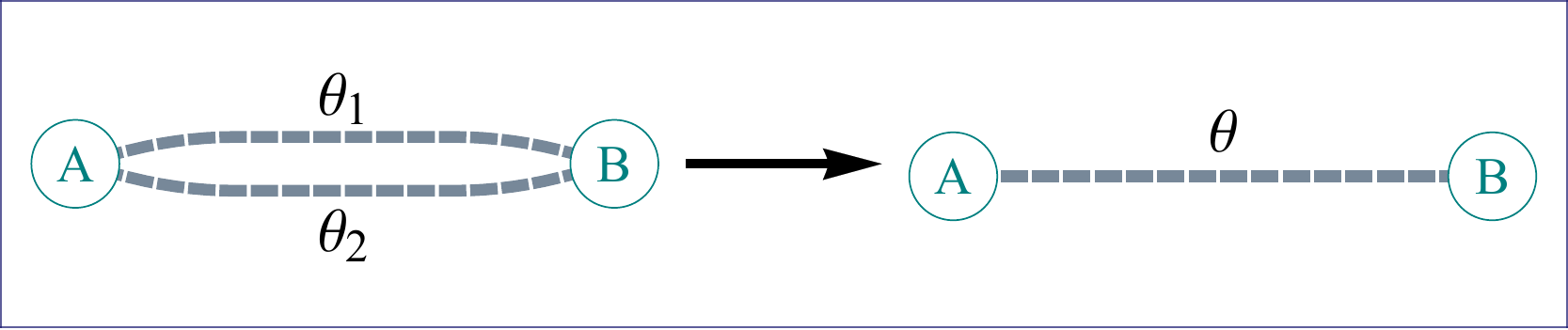}
		\subcaption{\label{fig_rules_b}}
		\vspace{5.7mm}
		
		\includegraphics[width=8.23cm]{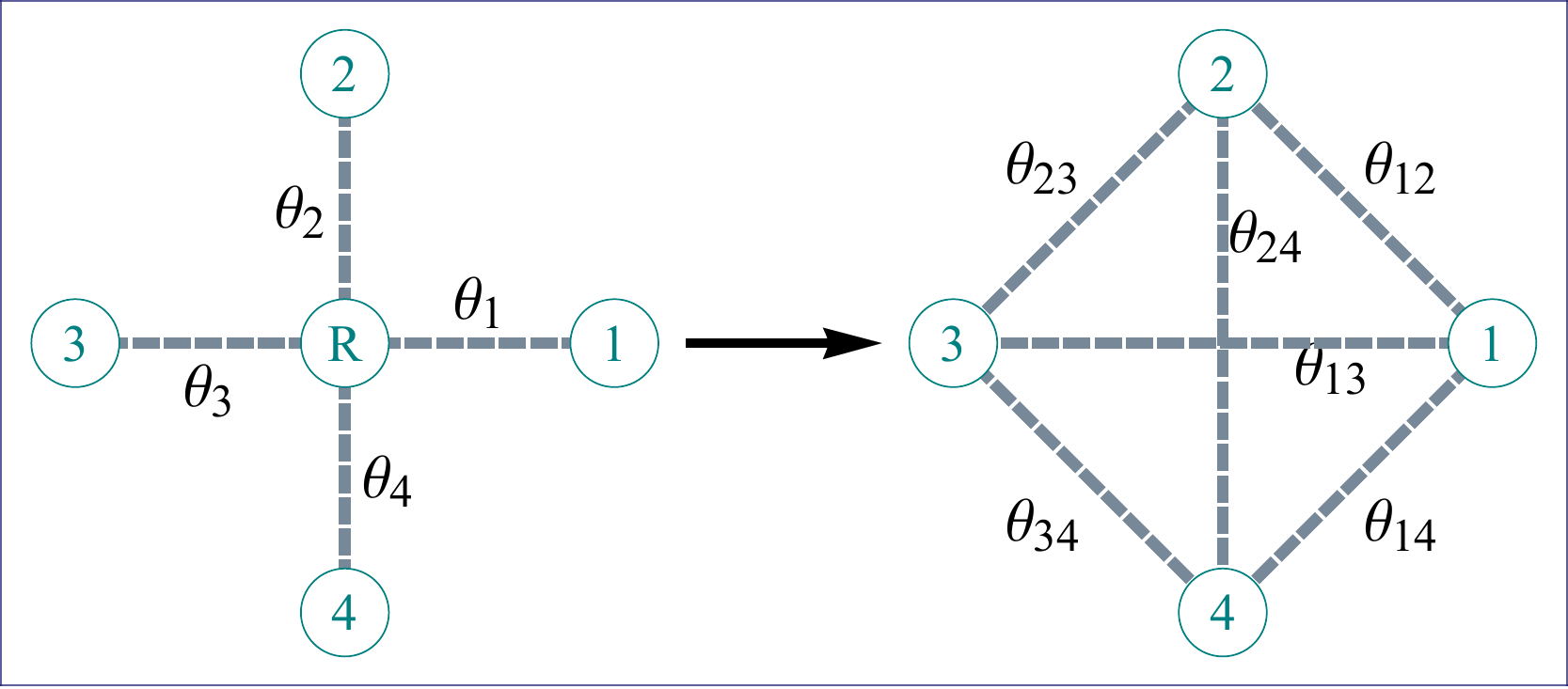}
		\subcaption{\label{fig_rules_c}}		
	\end{minipage}
	\begin{minipage}[b]{7.3cm}
		\centering
		
		\includegraphics[width=6.8cm]{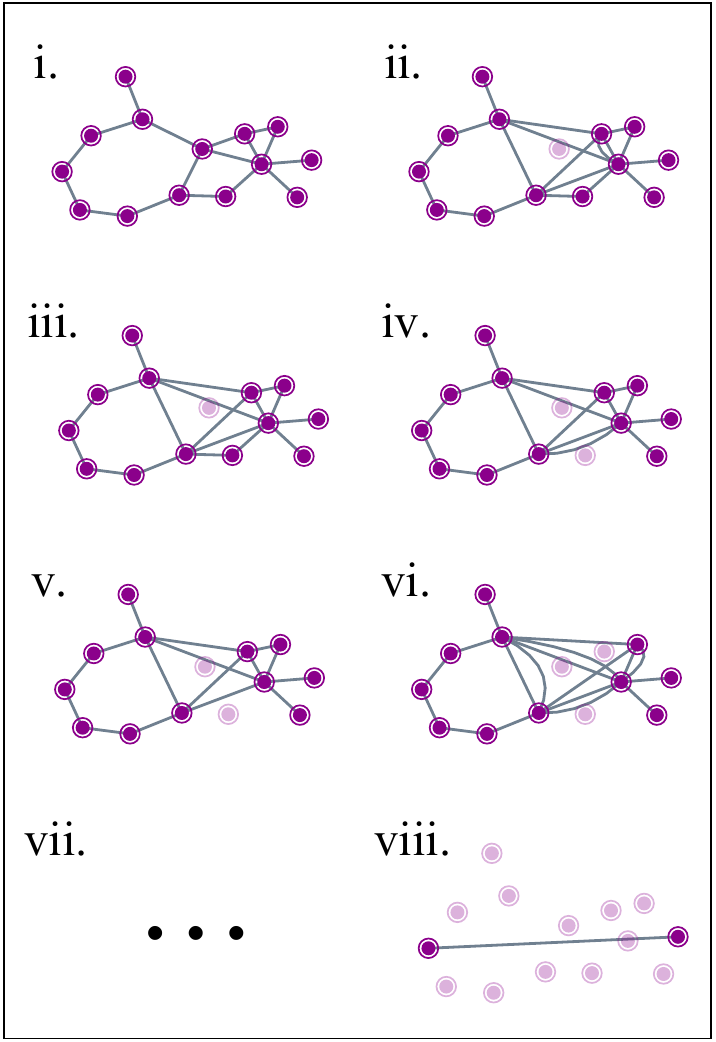}
		\subcaption{\label{fig_QN-transform}}
	\end{minipage}
	\caption{\label{fig_rules}{Connectivity rules and transform diagrams.}
		{\subref{fig_rules_a}} Series rule. {\subref{fig_rules_b}} Parallel rule. {\subref{fig_rules_c}} Star-mesh transform from an $n$-graph to an $(n-1)$-graph, solvable by applying series and parallel rules recursively through a group of $n(n-1)/2$ coupled equations. {\subref{fig_QN-transform}} Contracting an arbitrary QN to a pair of nodes using a sequence of star-mesh transforms (i.~$\to$~viii.).\hfill\hfill}
\end{figure}

A star-mesh transform~\cite{star-mesh_v70} can be built upon only series and parallel rules [Figs.~\ref{fig_rules_a} and~\ref{fig_rules_b}] but not higher-order rules to map an $n$-node star graph to an $(n-1)$-node complete graph [Fig.~\ref{fig_rules_c}], establishing a local equivalence (in terms of connectivity) between the two graphs. Note that each time the transform is applied, a node is degraded. Therefore, by applying the star-mesh transform consecutively, any network can be reduced to two nodes [Fig.~\ref{fig_QN-transform}]. The final weight $\theta$ of the link between the two nodes should well approximate the original connectivity between them, compensating our ignorance of the unknown higher-order rules. If we wish to look at the connectivity between not two nodes but two sets of nodes, e.g.,~two separate boundaries, then we can simply let all nodes from one set be always connected, which essentially behave just like one ``meta node''. Technically, this amounts to manually setting the link weights within the two sets to be always $\pi/4$.

We believe that our approximation is analogous to the real-space renormalization group (RG) for percolation theory: the original local structure of connectivity is replaced by an equivalent structure with less degrees of freedom. What is different, however, is that the star-mesh transform is more general and works for any kind of networks, not only lattices.

Just like there are different ways to do real-space RG on lattices, our method also allows different approximation procedures: one can degrade the nodes in the network in different orders. The results should not deviate too much from each other.

Details on how to solve the star-mesh transform are given below.

\subsection{Solve the star-mesh transform}

We define $\mathcal{G}(n;\theta_1,\theta_2,\cdots,\theta_n)$, denoted $\mathcal{G}(n)$ for simplicity, to be a star graph with one root vertex and $n$ leaf vertices. The weights of the $n$ edges are given, from $\theta_1$ to $\theta_n$. We define $\mathcal{G}'(n;\theta_{12},\theta_{13},\cdots,\theta_{1n},\cdots,\theta_{n-1,n})$, denoted $\mathcal{G}'(n)$, to be the star-mesh transform of $\mathcal{G}(n)$, i.e.,~a $n$-complete graph that has $n(n-1)/2$ edges with different weights.

The equivalence between $\mathcal{G}(n)$ and $\mathcal{G}'(n)$ are formatted as $n(n-1)/2$ independent equations,
\begin{eqnarray}
\label{star-mesh-eqns}
\text{seri}\left(\theta_1,\theta_2\right) &=&c\left(1,2;\mathcal{G}'\left(n\right)\right),\nonumber\\
\text{seri}\left(\theta_1,\theta_3\right) &=&c\left(1,3;\mathcal{G}'\left(n\right)\right),\nonumber\\
&\cdots,&\nonumber\\
\text{seri}\left(\theta_1,\theta_n\right) &=&c\left(1,n;\mathcal{G}'\left(n\right)\right),\nonumber\\
&\cdots,&\nonumber\\
\text{seri}\left(\theta_{n-1},\theta_n\right) &=&c\left(n-1,n;\mathcal{G}'\left(n\right)\right),
\end{eqnarray}
where $\text{seri}(\theta_i,\theta_j)$ is the series-sum of $\theta_i$ and $\theta_j$ based on the series rule, and the more complicated
$c(i,j;\mathcal{G}'(n))$ is the net weight between vertices $i$ and $j$ of the complete graph $\mathcal{G}'(n)$. We arbitrarily choose a vertex from $\mathcal{G}'(n)$ (w.l.o.g., the last one, $n$) to be the new root of a sub-star-graph of $\mathcal{G}'(n)$ constructed from the $n-1$ edges that connect the root to the other $n-1$ vertices. We transform this sub-star-graph $(\text{sub}\mathcal{G}')(n-1)$ into a $(n-1)$-complete graph, denoted by $(\text{sub}\mathcal{G}')'(n-1)$, and combine it with what is left untransformed, {$\mathcal{G}'(n)\setminus\nobreak(\text{sub}\mathcal{G}')(n-\nobreak1)$}, which is also a $(n-1)$-complete graph. %We format the procedure and write the resulting graph to be $\text{Comb}\left((\text{sub}\mathcal{G}')'(n-1),\mathcal{G}'(n)\setminus(\text{sub}\mathcal{G}')(n-1)\right)$. Here, 
We define $\text{Comb}\left(\mathcal{G}_{\alpha},\mathcal{G}_{\beta}\right)$ as the new graph derived by setting each edge weight to be $\theta_{ij}=\text{para}(\alpha_{ij},\beta_{ij})$,  which is the parallel-sum of $\alpha_{ij}\in \mathcal{G}_{\alpha} $ and $\beta_{ij}\in\mathcal{G}_{\beta}$ based on the parallel rule. We can calculate $c(i,j;\mathcal{G}'(n))$ by first solving a $(n-1)$-complete graph now,
\begin{equation}
\label{c-recur}
c(i,j;\mathcal{G}'(n))=c(i,j;\text{Comb}\left((\text{sub}\mathcal{G}')'(n-1),\mathcal{G}'(n)\setminus(\text{sub}\mathcal{G}')(n-1)\right)).
\end{equation}
Thus $c(i,j;\mathcal{G}'(n))$ is calculable through recursions. %The star-mesh transform is solvable because it needs only a proper definition of the series and parallel rules.
Note that, because Eq.~(\ref{c-recur}) also involves a $(n-1)$-level star-mesh transform, the entire procedure is a double recursion, the cost growing faster than exponential.

Equations~(\ref{star-mesh-eqns}) have closed-form solutions for calculating the net resistance in a resistance network. In contrast, we have not found any closed-form solution for our concurrence percolation theory (ConPT) and thus have to use Broyden's root-finding algorithm to numerically find the $n(n-1)/2$ weights $\theta_{ij}$ that satisfy Eq.~(\ref{star-mesh-eqns}). Broyden's algorithm does not require a calculation of the accurate Jacobian matrix at each iteration but updates the Jacobian using partial results from the previous iteration. This allows a reduction in cost, especially when the equations are defined recursively.  %However there are other techniques that can be applied to the algorithm, e.g.,~dynamic programming and faster initializations. Note that there are no guaranteed solutions, even though the system may look similar to known theories.
In practice, the recursive computation is carried out by symbolic expressions in \emph{Mathematica}. The performance threshold is $n \sim 11$ nodes which corresponds to solving $55$ double-recursive independent equations. Although proving the existence of solutions in Eq.~(\ref{star-mesh-eqns}) is challenging, our calculations show that the solutions are within a sufficiently small error range. 

Finally, the star-mesh transform is applied one after the other on all but two nodes in the network of our interest. Randomness is added to initialization and to the procedure of choosing in which order to degrade the nodes. We executed more than seven runs for each graph and each assigned weight, and then took the average to reduce the algorithm inaccuracy.

Usually, a more natural way to study percolation phenomena without using series and parallel rules or burdensome star-mesh transforms would be to apply a Monte Carlo method. One rudimentary application is the probabilistic simulation on clusters in classical percolation theory. There are also Monte Carlo methods applicable for resistance networks. We expect that a Monte Carlo method to ConPT, if it exists, should follow a similar procedure and significantly accelerate the calculation.

\newpage

\begin{figure}[h!]
	\centering
	\begin{minipage}[t]{51mm}
		\centering
		%\vspace{12mm}
		\includegraphics[width=51mm]{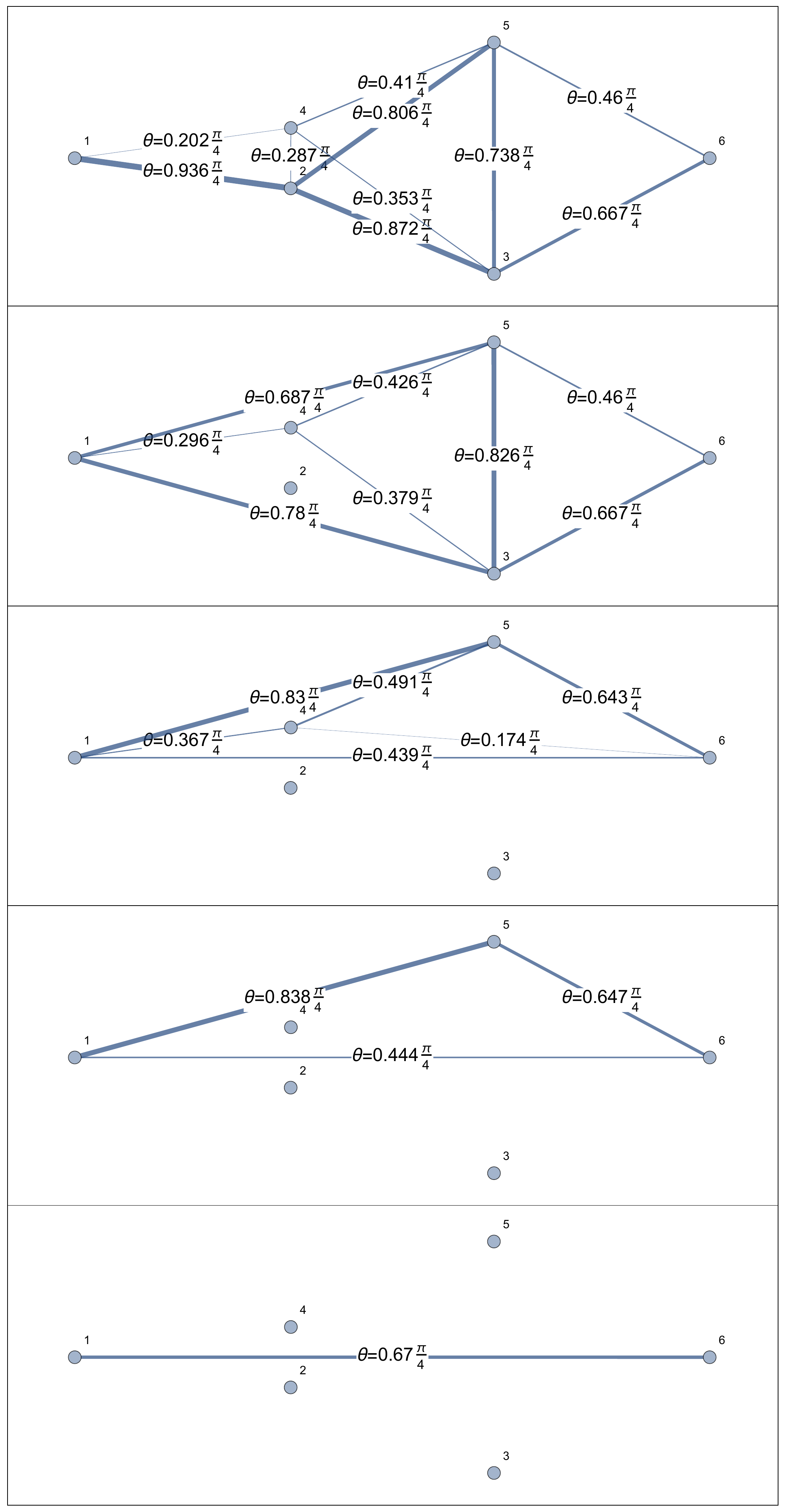}
		\subcaption{\label{fig_p2345}Classical. $2\to3\to4\to5$.}
		\includegraphics[width=51mm]{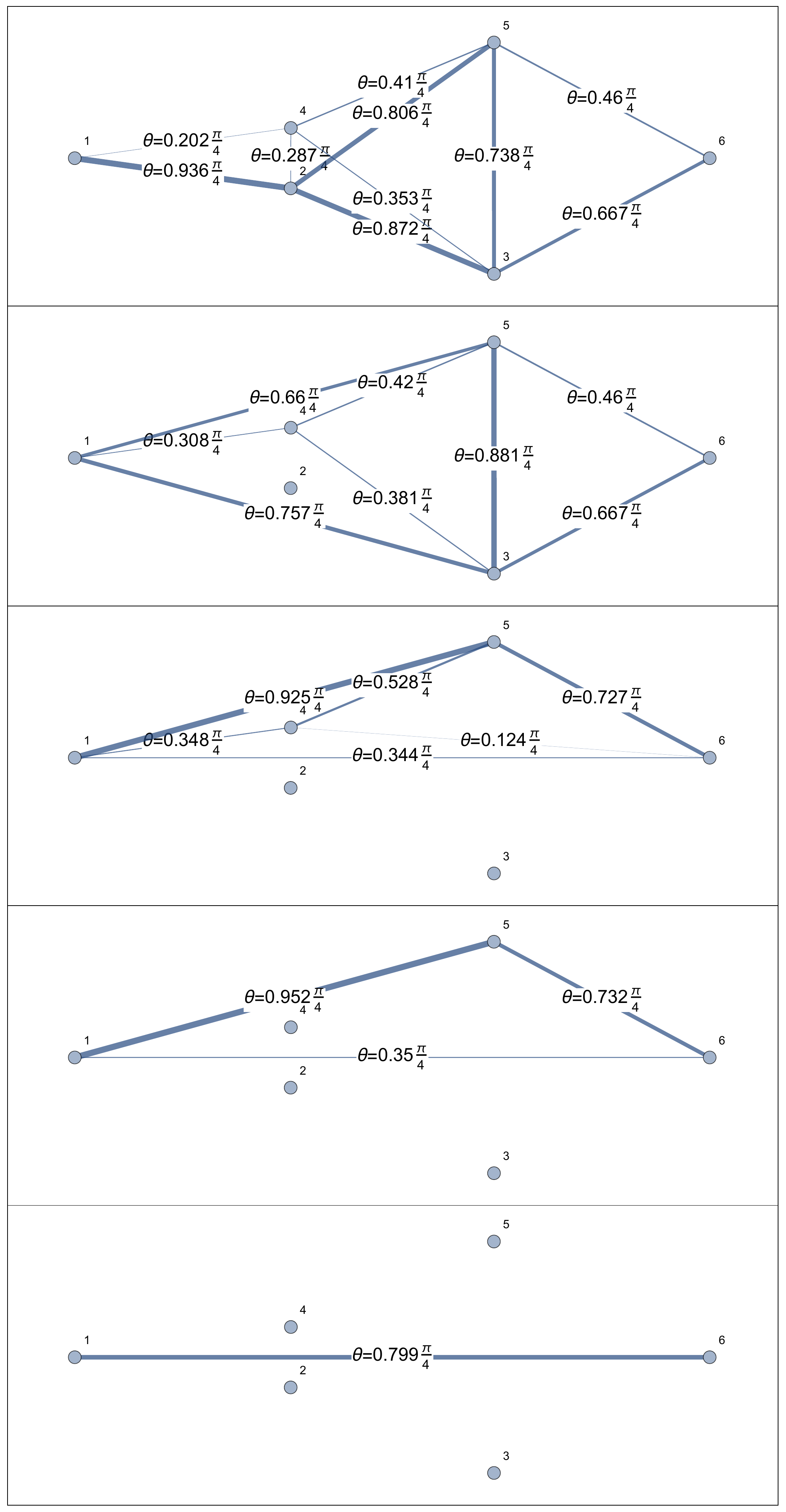}
		\subcaption{\label{fig_c2345}ConPT. $2\to3\to4\to5$.}
	\end{minipage}
	\begin{minipage}[t]{51mm}
		\centering
		%\vspace{12mm}
		\includegraphics[width=51mm]{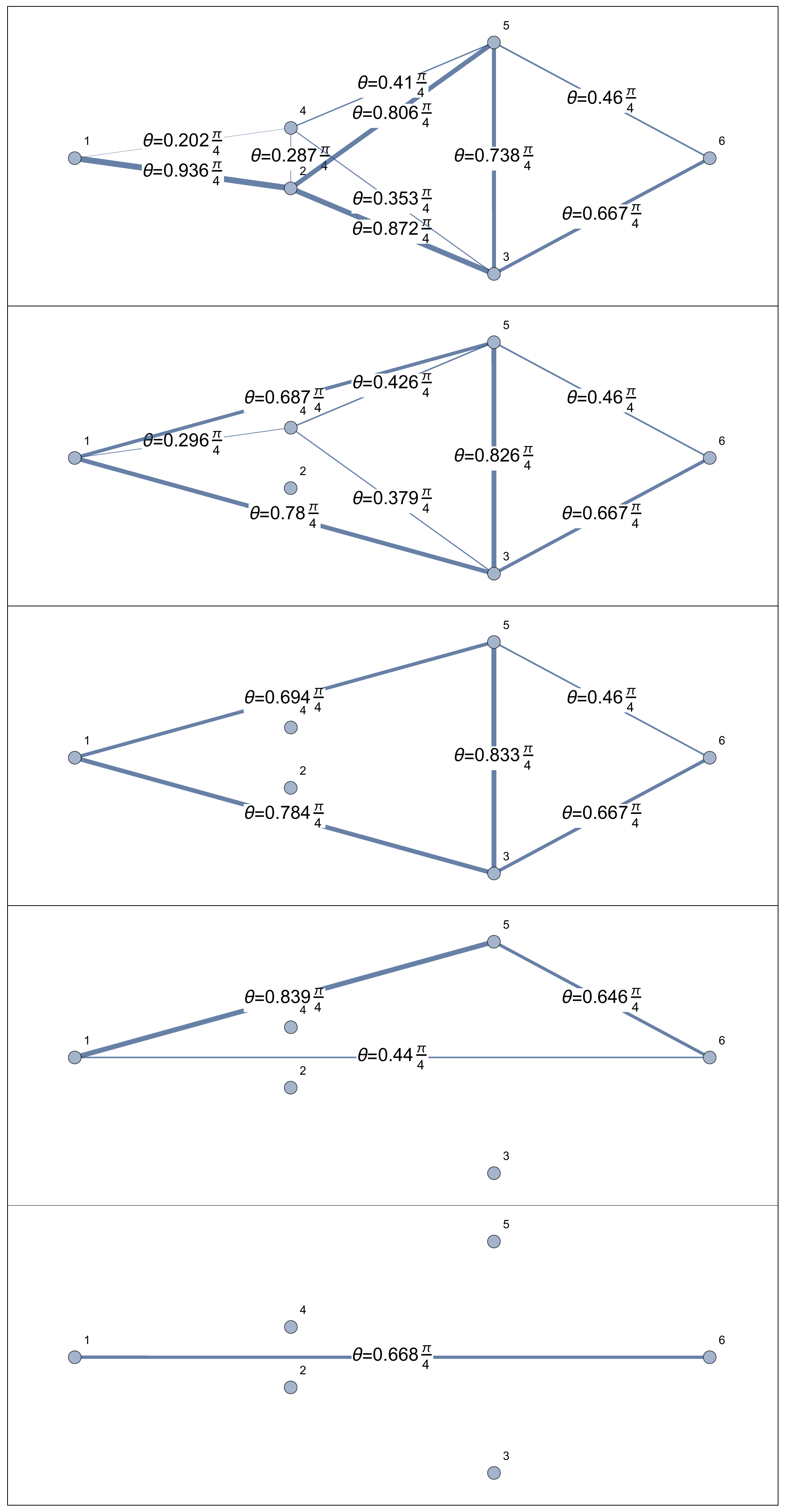}
		\subcaption{\label{fig_p2435}Classical. $2\to4\to3\to5$.}
		\includegraphics[width=51mm]{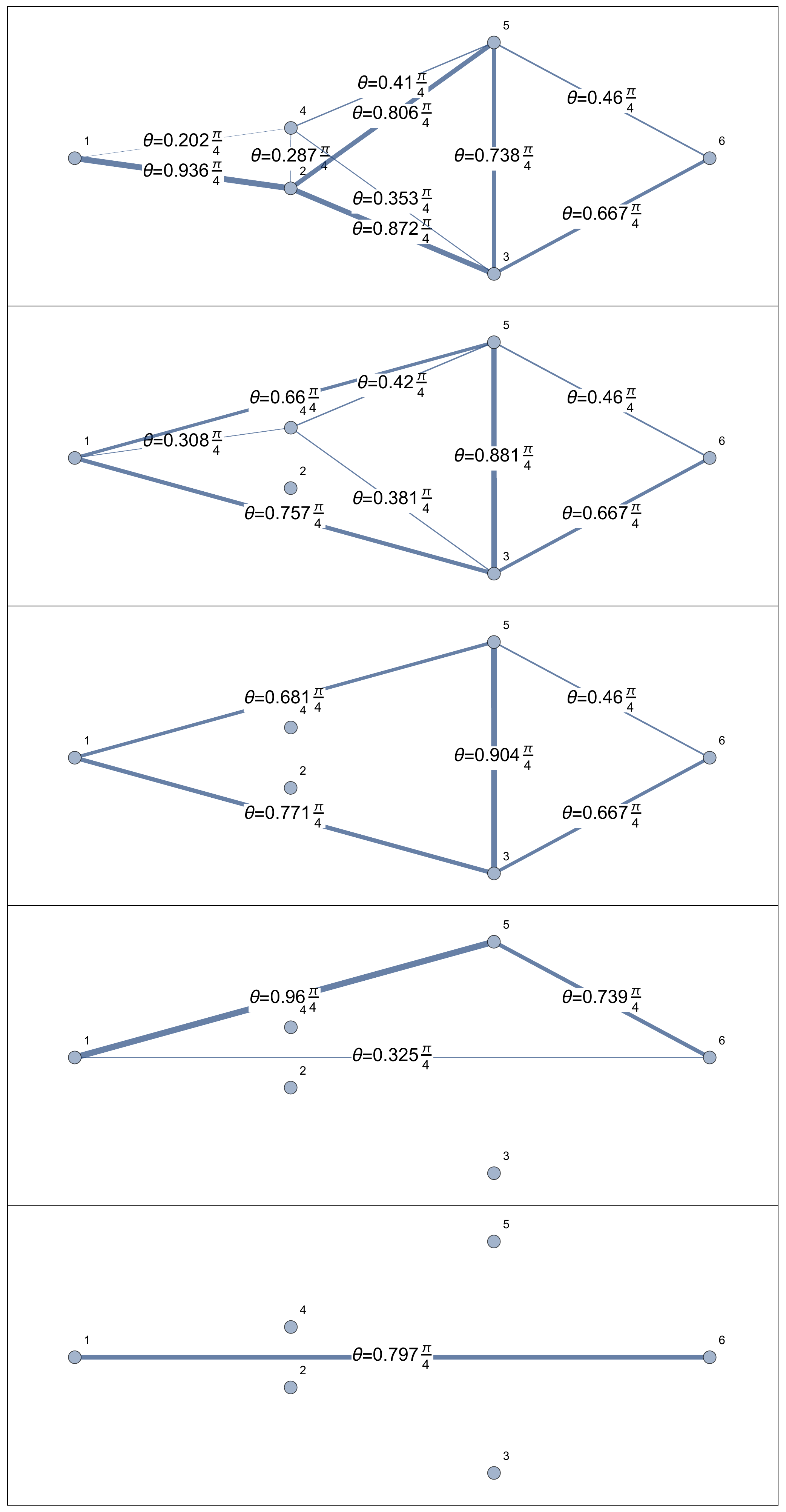}
		\subcaption{\label{fig_c2435}ConPT. $2\to4\to3\to5$.}
	\end{minipage}
	\begin{minipage}[t]{51mm}
		\centering
		%\vspace{12mm}
		\includegraphics[width=51mm]{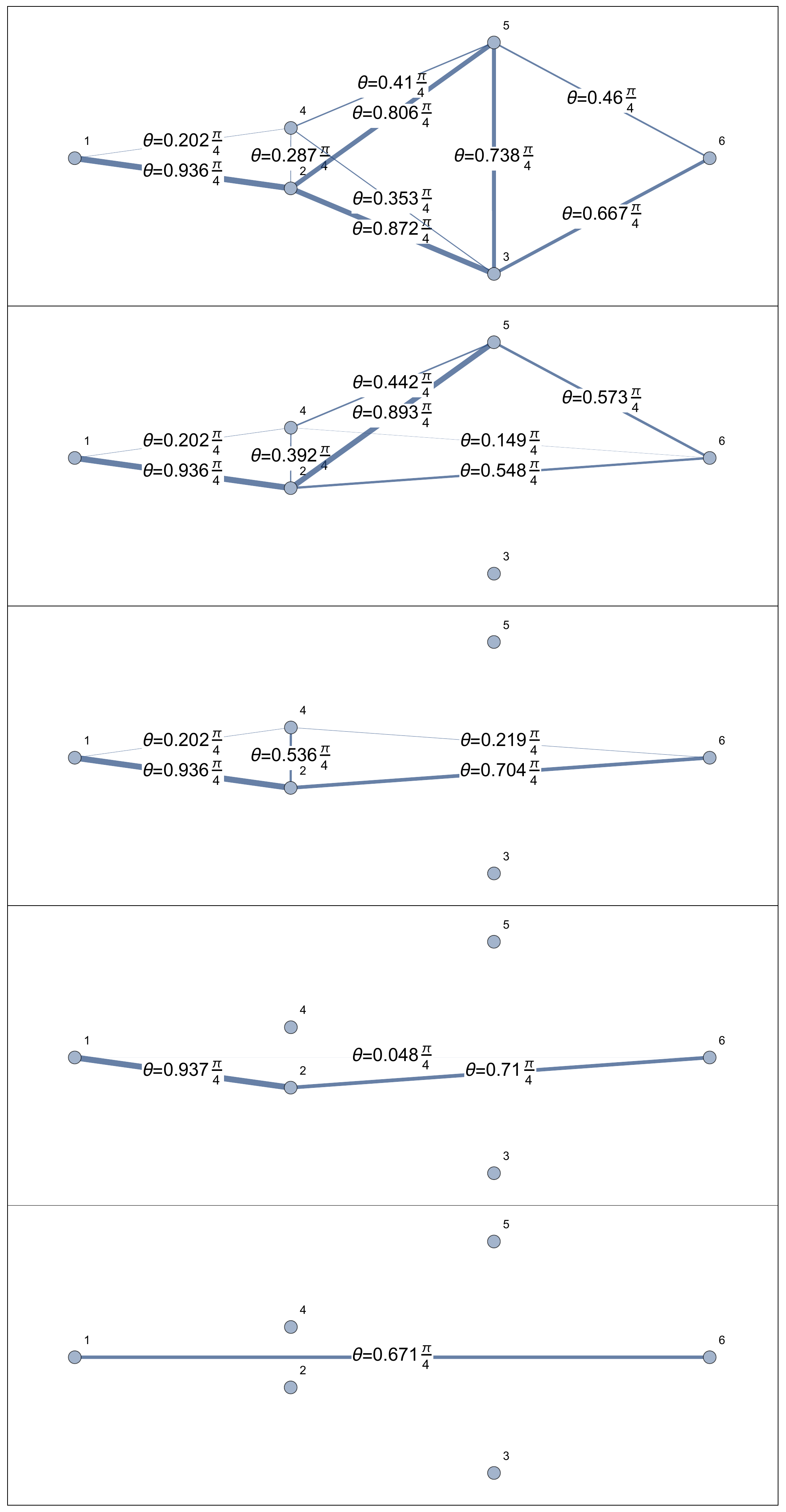}
		\subcaption{\label{fig_p3542}Classical. $3\to5\to4\to2$.}
		\includegraphics[width=51mm]{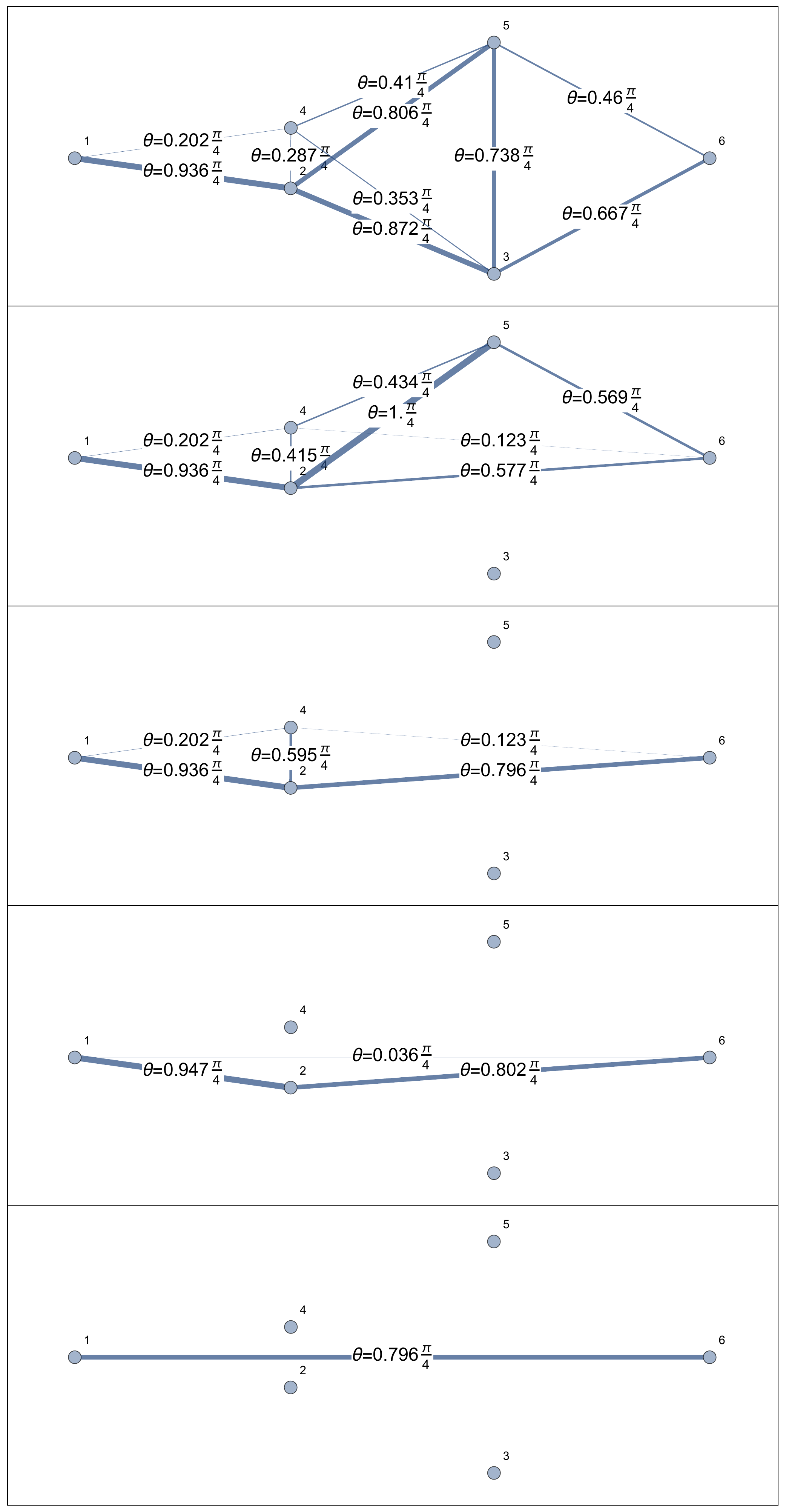}
		\subcaption{\label{fig_c3542}ConPT. $3\to5\to4\to2$.}
	\end{minipage}
	\caption{}
\end{figure}

\begin{figure}[h!]
	\ContinuedFloat
	\caption{\label{fig_compare}Two different percolation theories, classical percolation and ConPT, are used for studying entanglement transmission in a heterogeneous QN. Classical percolation theory predicts the final singlet conversion probability (SCP)~\cite{QEP_acl07}, i.e.,~the probability to establish a singlet between nodes~$1$~and~$6$. ConPT predicts the average concurrence of the final states to be established between the same two nodes. Different approximation procedures of the star-mesh transform are studied. \subref{fig_p2345},~\subref{fig_p2435},~and~\subref{fig_p3542}~For classical percolation theory, given $p_i\equiv2\sin^{2}\theta_i$ per link, the probability that at least one path connects nodes $1$ and $6$ is known exactly to be $P_\text{SC}\approx0.5016$ (obtained by $5\times10^4$ Monte Carlo simulations) which corresponds to $\theta_\text{SC}=\sin^{-1}\sqrt{P_\text{SC}/2}\approx0.67\pi/4$. Here, only series and parallel rules plus the star-mesh transform are used instead, and for three different approximation procedures the results of $\theta_\text{SC}$ are all consistent with the exact simulation result. The SCP between nodes $1$ and $6$ predicted by classical percolation theory is thus $\approx0.5016$.
	{\protect\subref{fig_c2345}},~{\protect\subref{fig_c2435}},~and~{\protect\subref{fig_c3542}}~The final concurrence between nodes $1$ and $6$ predicted by ConPT is $C_\text{SC}=\sin2\theta_\text{SC}\approx0.95$, given that $\theta_\text{SC}\approx0.80\pi/4$ for all three procedures. \hfill\hfill}
\end{figure}

\subsection{Example: entanglement transmission in a heterogeneous QN}

Figure~\ref{fig_compare} shows an example of how to calculate the entanglement transmission between nodes~$1$ and $6$ in a small heterogeneous QN using the star-mesh transform. Different approximation procedures (i.e.,~degrading the other nodes in different orders) are studied.

By comparing the three approximate results [Figs.~\ref{fig_p2345},~\ref{fig_p2435},~and~\ref{fig_p3542}] with the exact Monte Carlo simulation result ($P_\text{SC}\approx0.5016$) for classical percolation theory, we see that the star-mesh transform indeed works very well as an RG-like approximation. We thus hypothesize that this approximation should equally work well for ConPT. We see again that the three approximate results for ConPT [Figs.~\ref{fig_c2345},~\ref{fig_c2435},~and~\ref{fig_c3542}]   do not differ much from each other but are notably higher than the classical ones (which shows the quantum advantage of ConPT over classical percolation theory).

\subsection{Consistence in different star-mesh approximation procedures}

\begin{figure}[h!]
	\centering
	\begin{minipage}[t]{60mm}
		\centering
		%\vspace{12mm}
		\includegraphics[width=60mm]{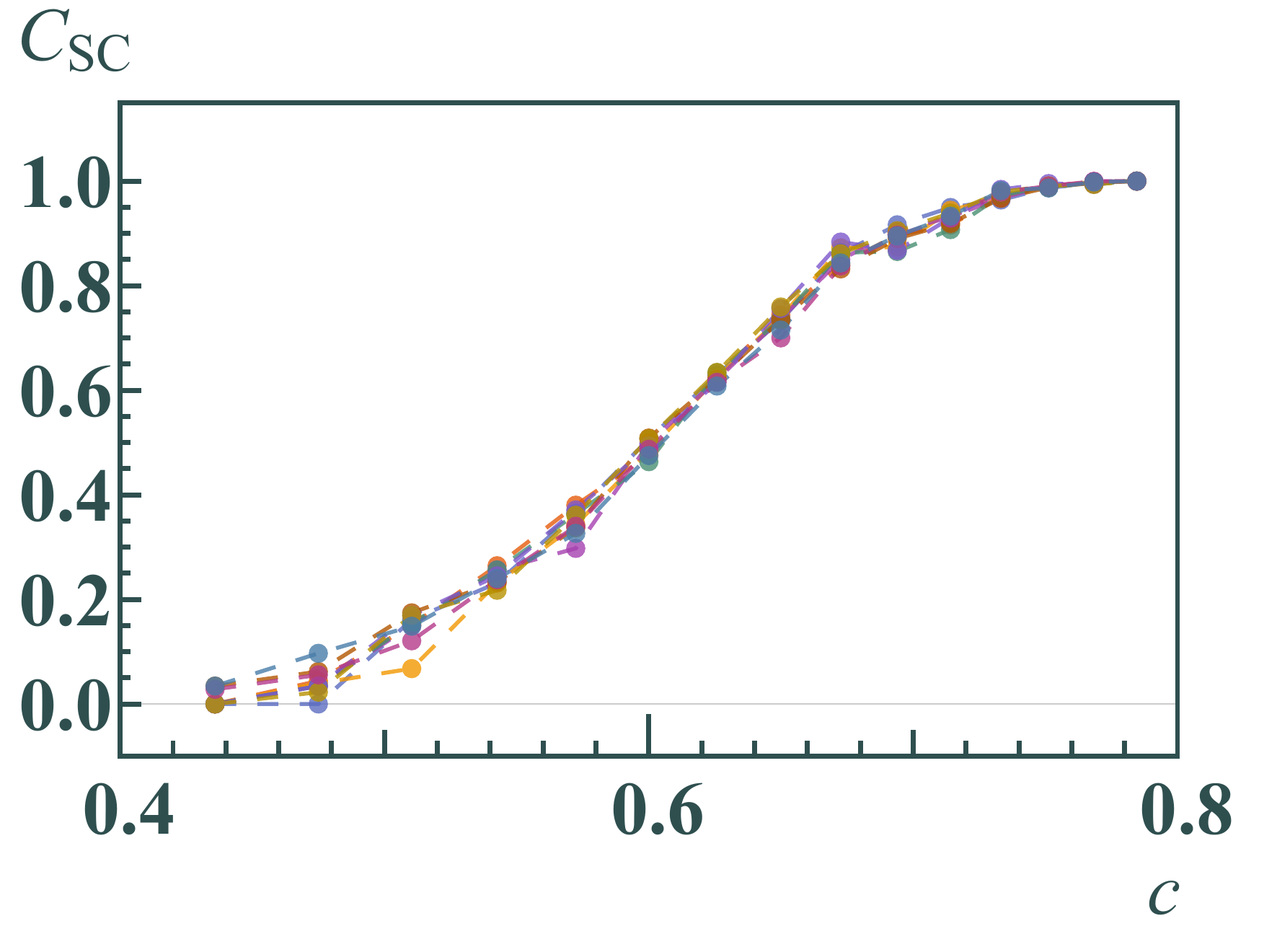}
		\subcaption{\label{fig_l5}}
		\includegraphics[width=60mm]{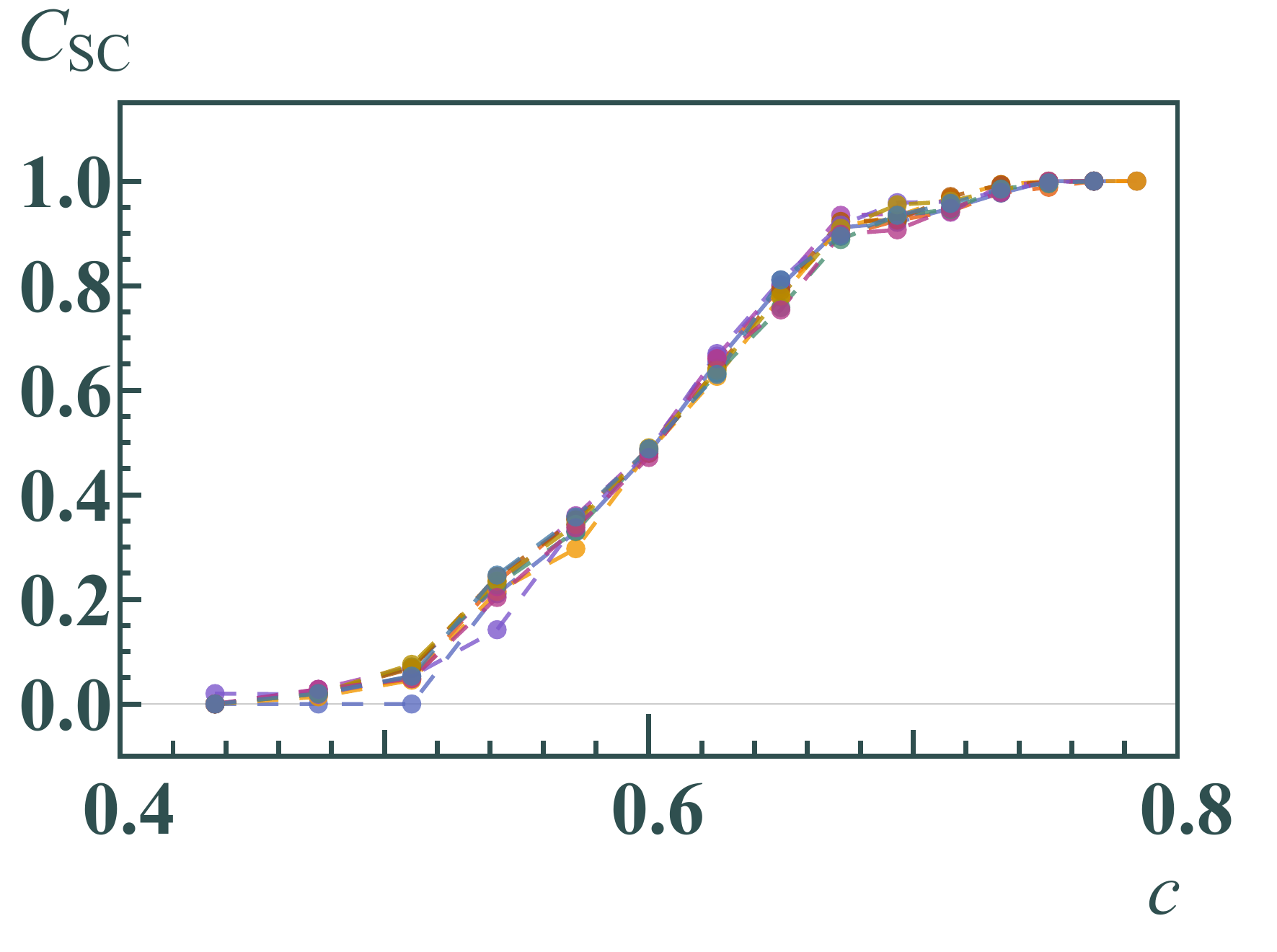}
		\subcaption{\label{fig_l6}}
	\end{minipage}
	\begin{minipage}[t]{60mm}
		\centering
		%\vspace{12mm}
		\includegraphics[width=60mm]{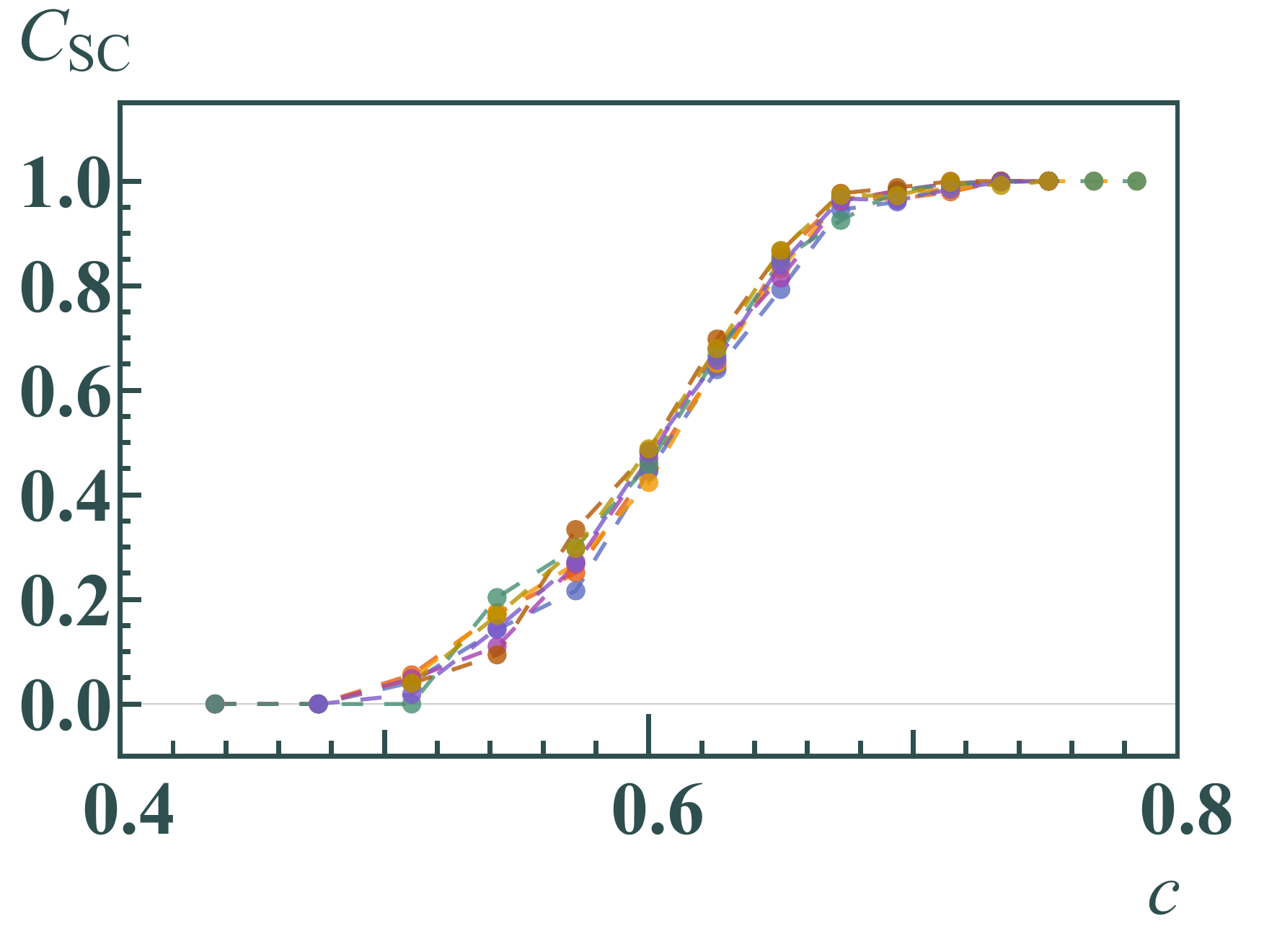}
		\subcaption{\label{fig_l7}}
		\includegraphics[width=60mm]{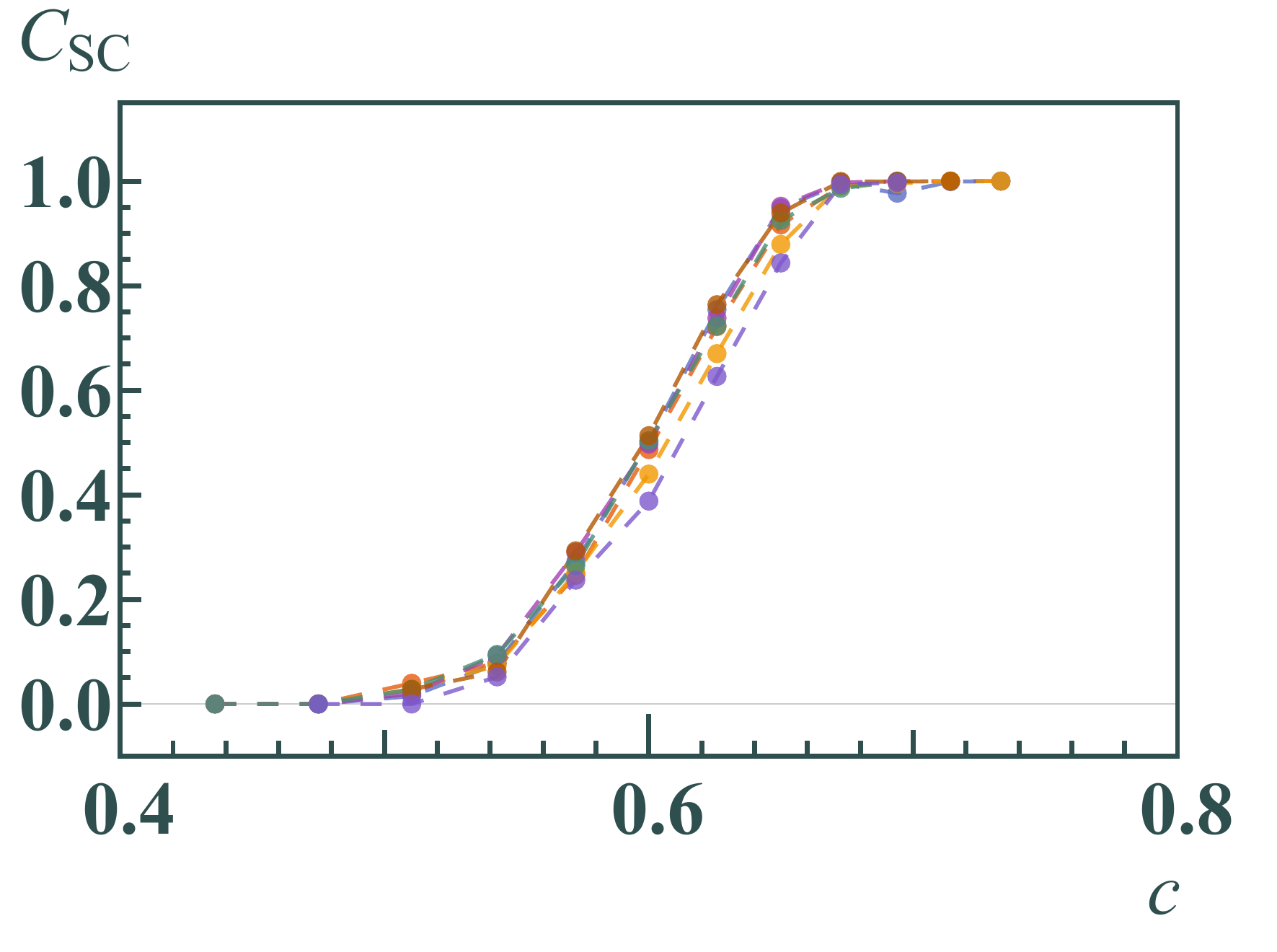}
		\subcaption{\label{fig_l8}}
	\end{minipage}
	\caption{\label{fig_diff-procedure}The star-mesh transform is performed for ConPT on a 2D square lattice of size \subref{fig_l5} $L=5$, \subref{fig_l6} $L=6$, \subref{fig_l7} $L=7$, and \subref{fig_l8} $L=8$. Different colors denote different approximation procedures, meaning the nodes are degraded in different random orders.\hfill\hfill}
\end{figure}

Figure~\ref{fig_diff-procedure} shows different approximation procedures (i.e.,~degrading the nodes in different orders) applied on the 2D square lattice for calculating ConPT. We can see that the difference between the procedures is rather small, and thus the star-mesh approximation is consistent. In our paper, we took the average before estimating the ConPT threshold.
\newpage

\newpage
\section{The Bethe Lattice}

\begin{figure}[t!]
	\centering
	\begin{minipage}[t]{38mm}
		\centering
		{\includegraphics[height=31mm]{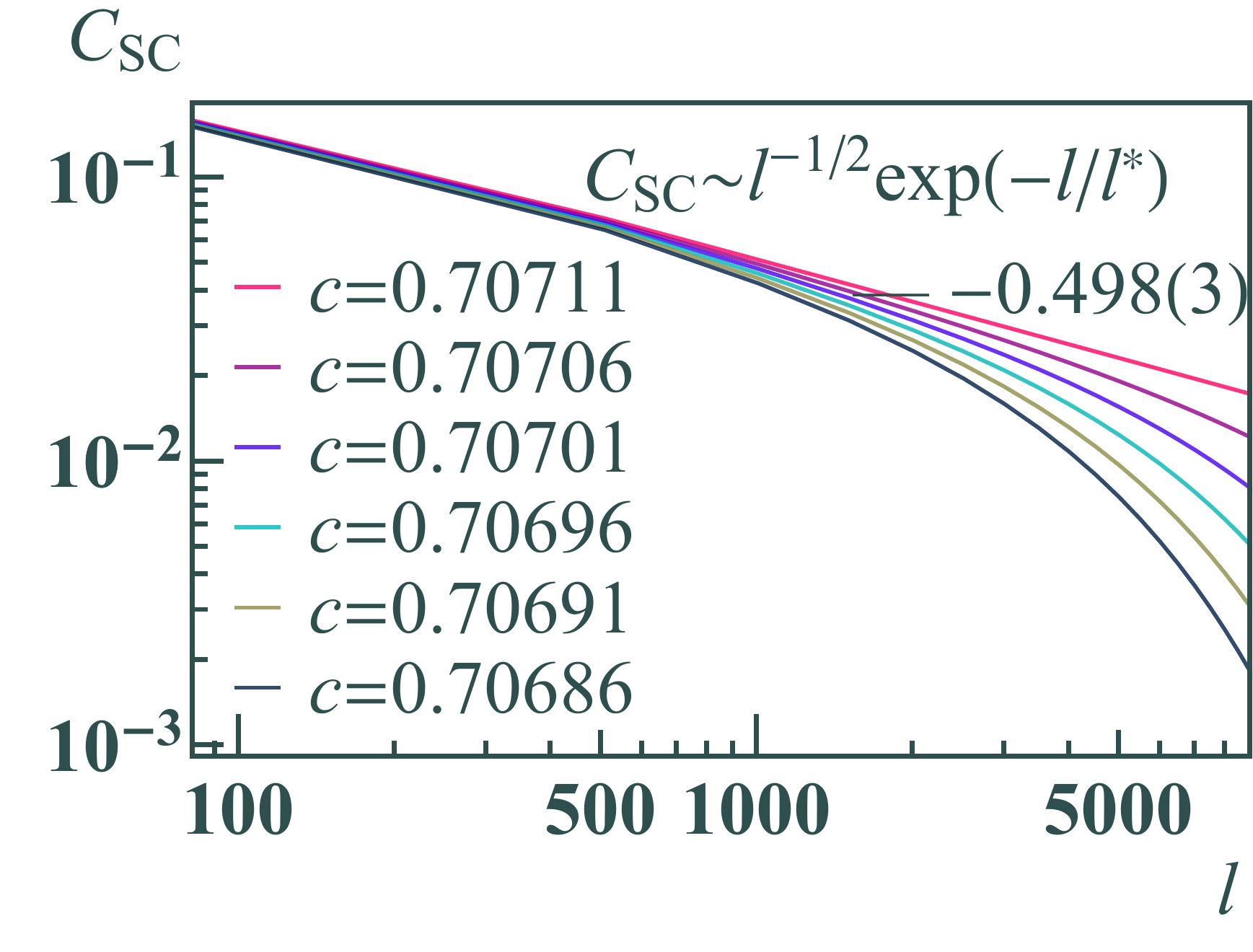}
			\vspace{-9mm}
			\subcaption{\label{fig_bethe_scaling_c10}}}
	\end{minipage}
	\begin{minipage}[t]{38mm}
		\centering
		{\includegraphics[height=31mm]{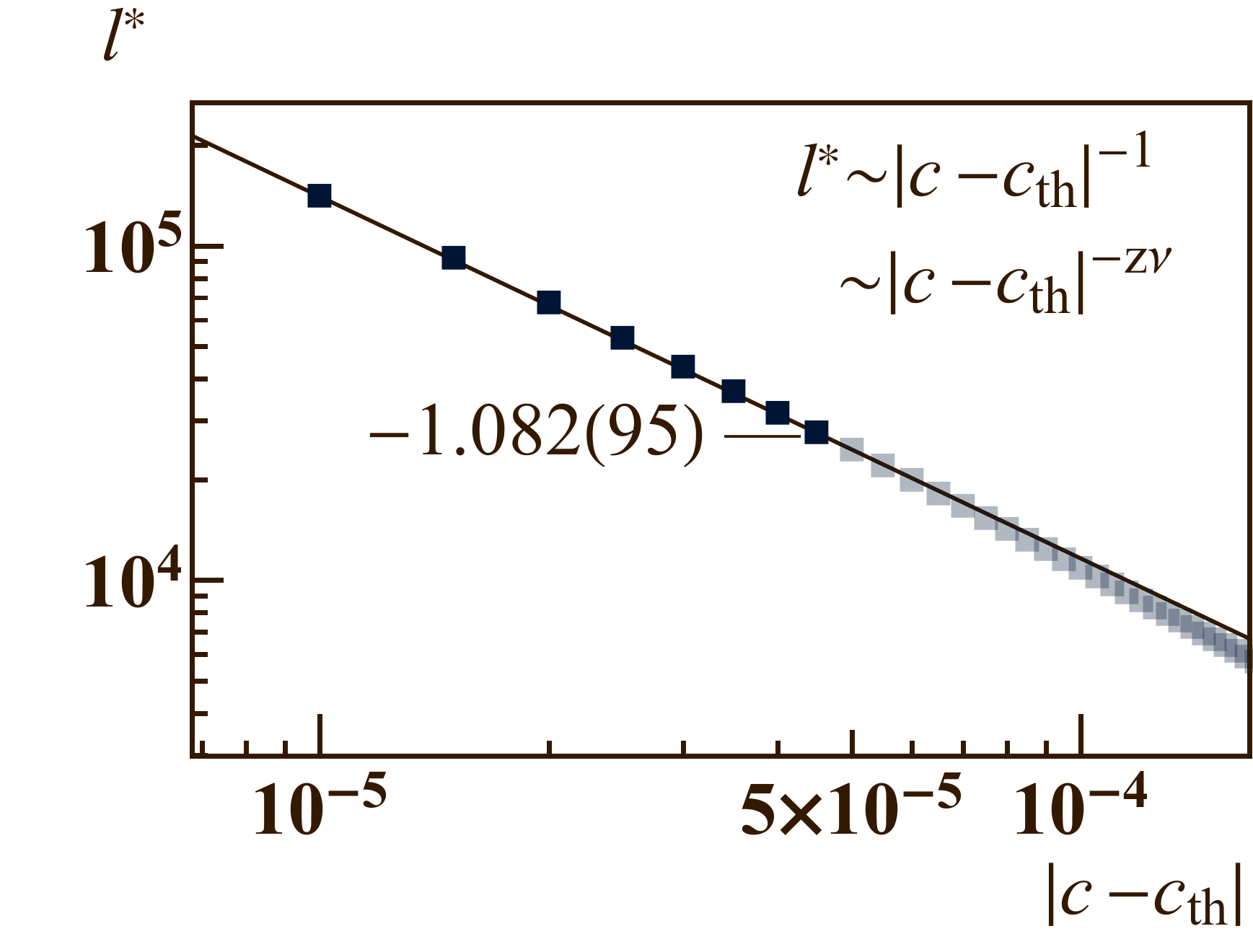}
			\vspace{-9mm}
			\subcaption{\label{fig_bethe_scaling_c11s}}}
	\end{minipage}
	\addtocounter{subfigure}{3}
	\begin{minipage}[t]{38mm}
		\centering
		{\includegraphics[height=31mm]{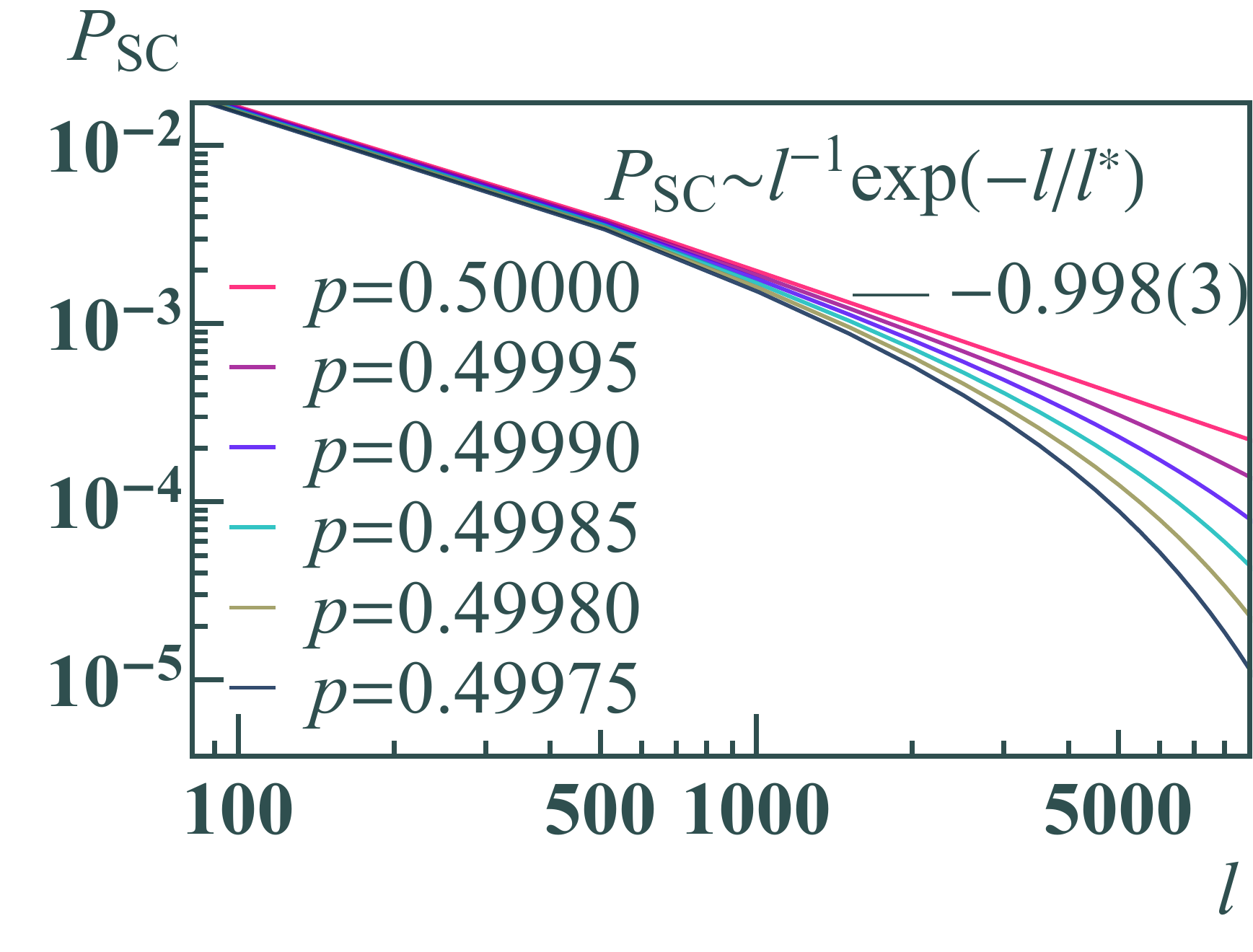}
			\vspace{-9mm}
			\subcaption{\label{fig_bethe_scaling_p10}}}
	\end{minipage}
	\begin{minipage}[t]{38mm}
		\centering
		{\includegraphics[height=31mm]{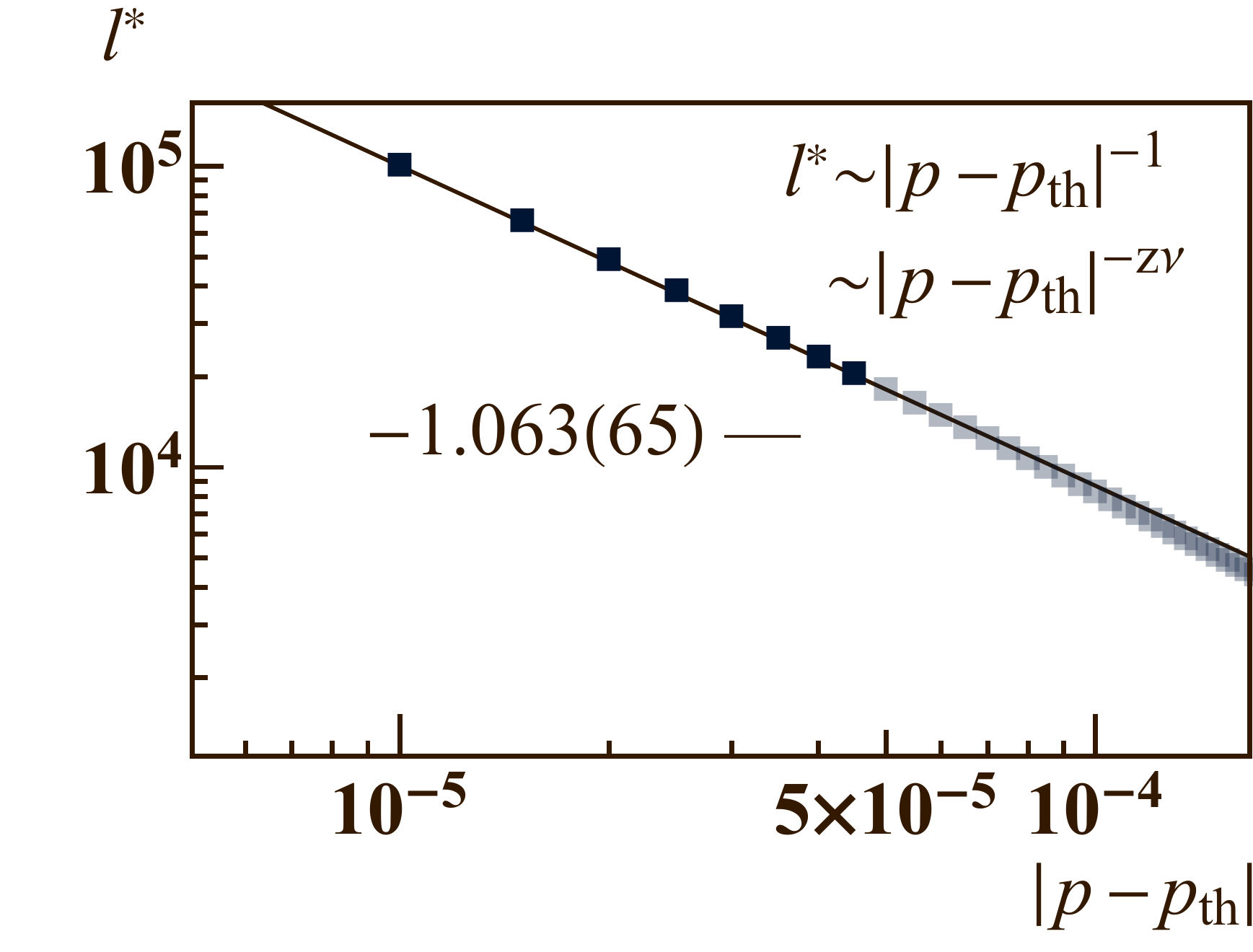}
			\vspace{-9mm}
			\subcaption{\label{fig_bethe_scaling_p11}}}
	\end{minipage}
	
	\addtocounter{subfigure}{-5}
	\begin{minipage}[t]{38mm}
		\centering
		{\includegraphics[height=31mm]{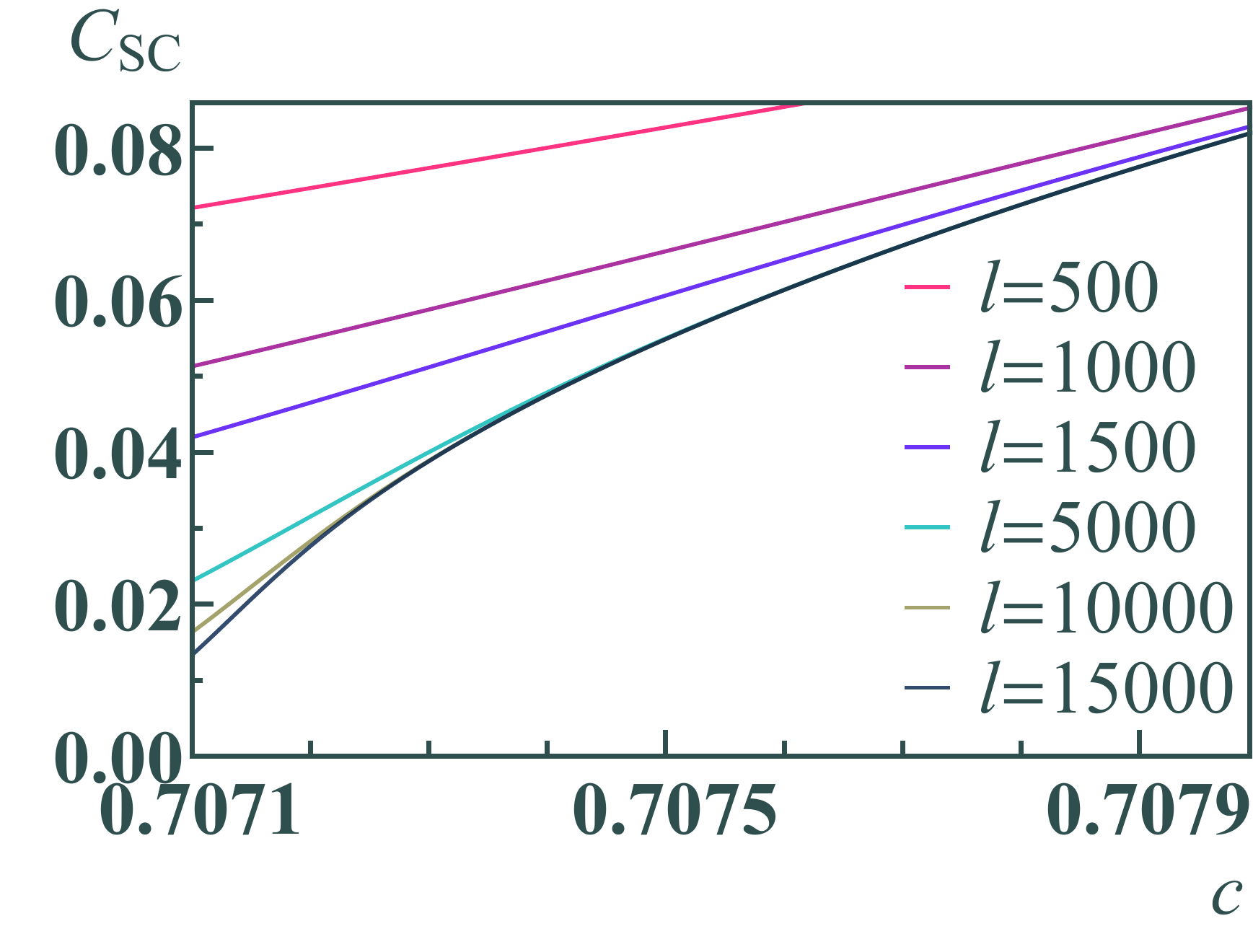}
			\vspace{-9mm}
			\subcaption{\label{fig_bethe_scaling_c20}}}
	\end{minipage}
	\begin{minipage}[t]{38mm}
		\centering
		{\includegraphics[height=31mm]{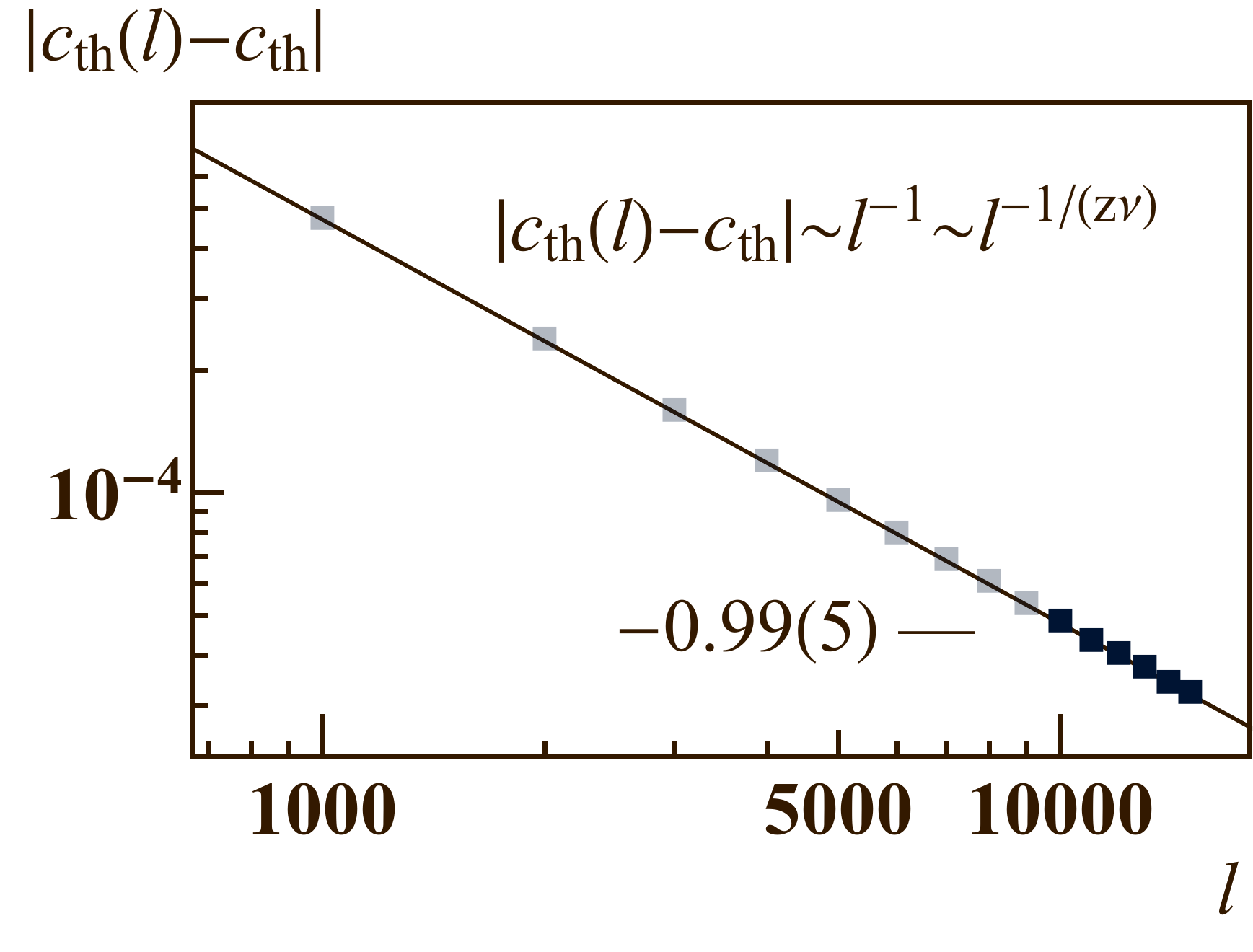}
			\vspace{-9mm}
			\subcaption{\label{fig_bethe_scaling_c21s}}}
	\end{minipage}
	\addtocounter{subfigure}{3}
	\begin{minipage}[t]{38mm}
		\centering
		{\includegraphics[height=31mm]{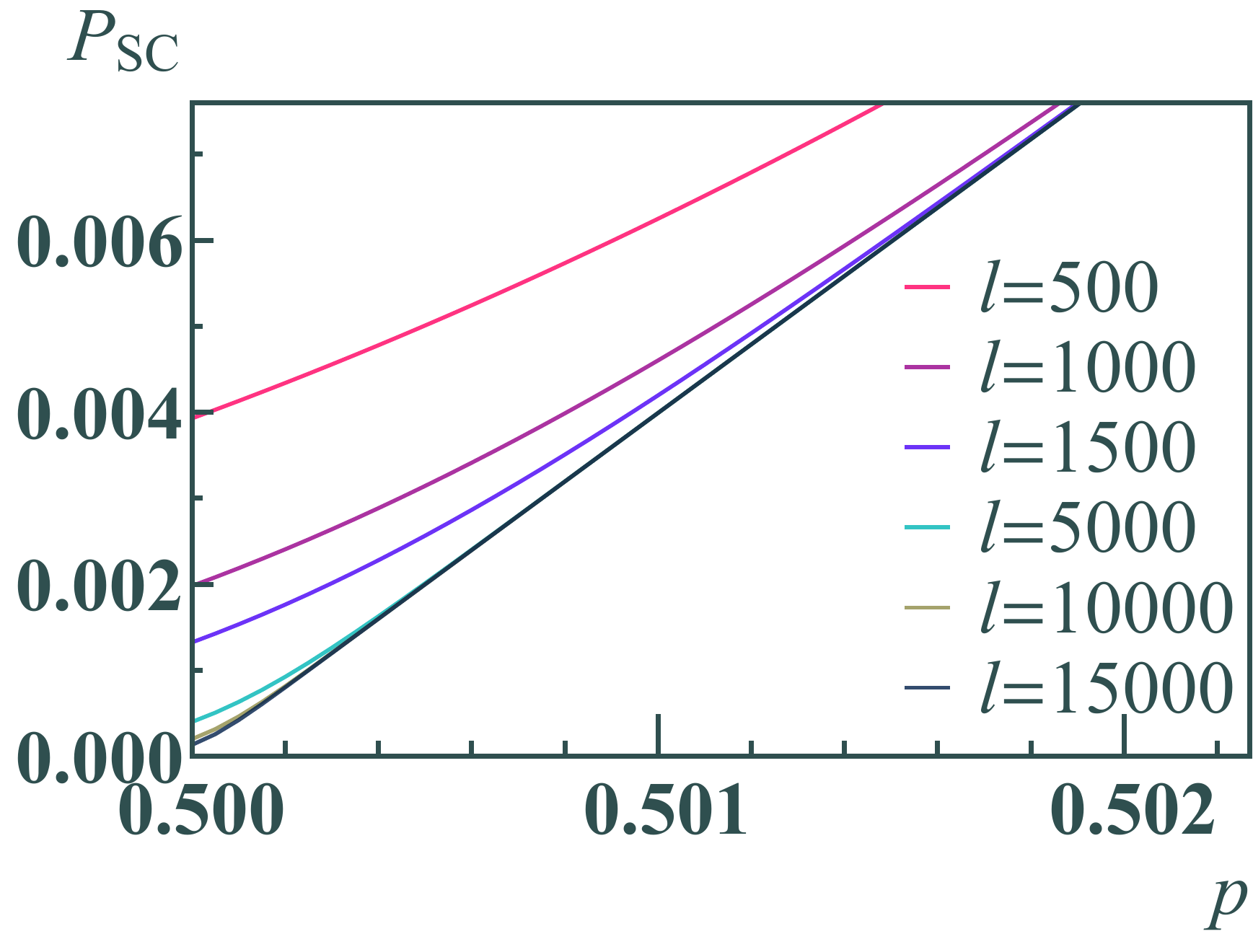}
			\vspace{-9mm}
			\subcaption{\label{fig_bethe_scaling_p20}}}
	\end{minipage}
	\begin{minipage}[t]{38mm}
		\centering
		{\includegraphics[height=31mm]{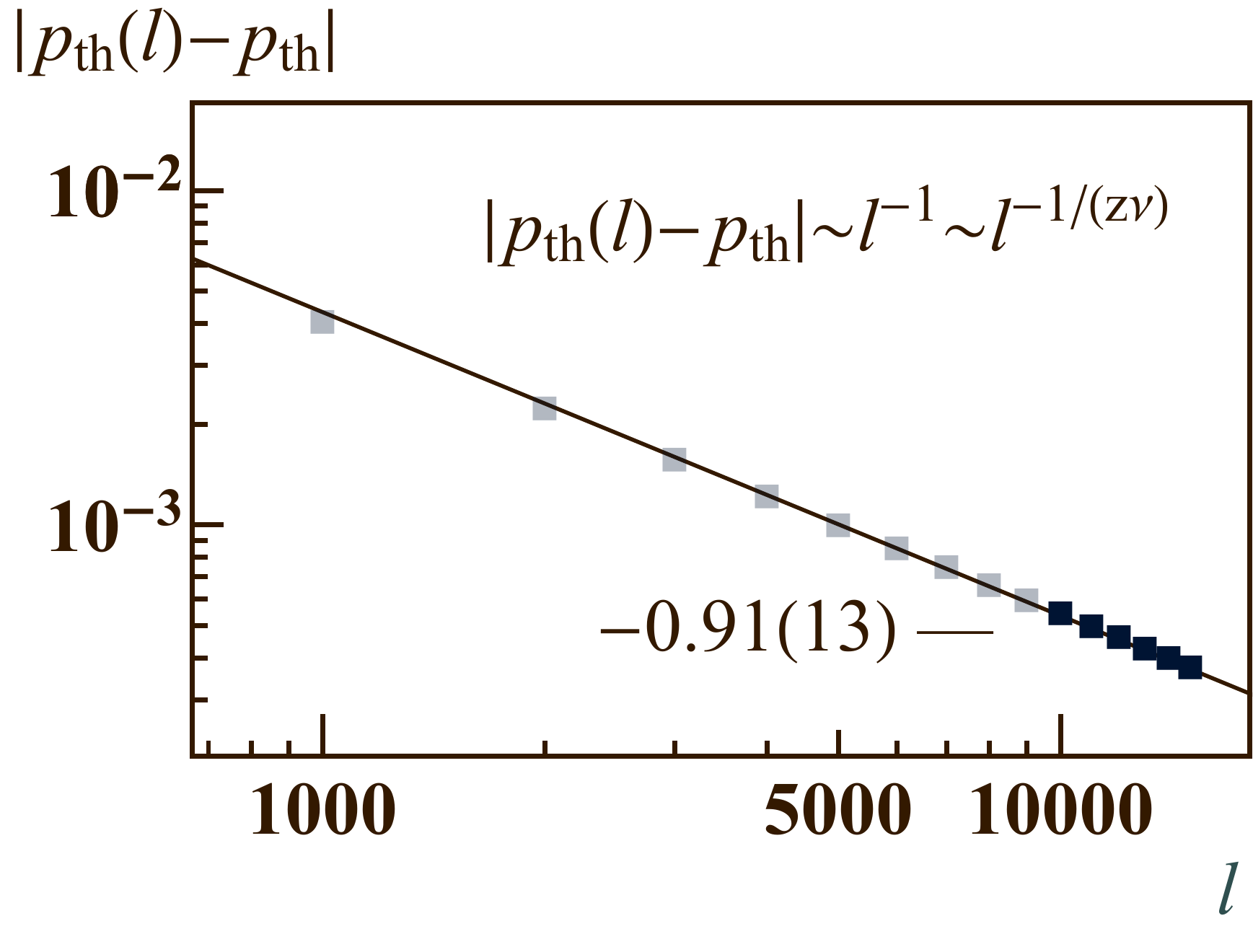}
			\vspace{-9mm}
			\subcaption{\label{fig_bethe_scaling_p21}}}
	\end{minipage}
	
	\addtocounter{subfigure}{-5}
	\begin{minipage}[t]{70mm}
		\centering
		\includegraphics[height=31mm]{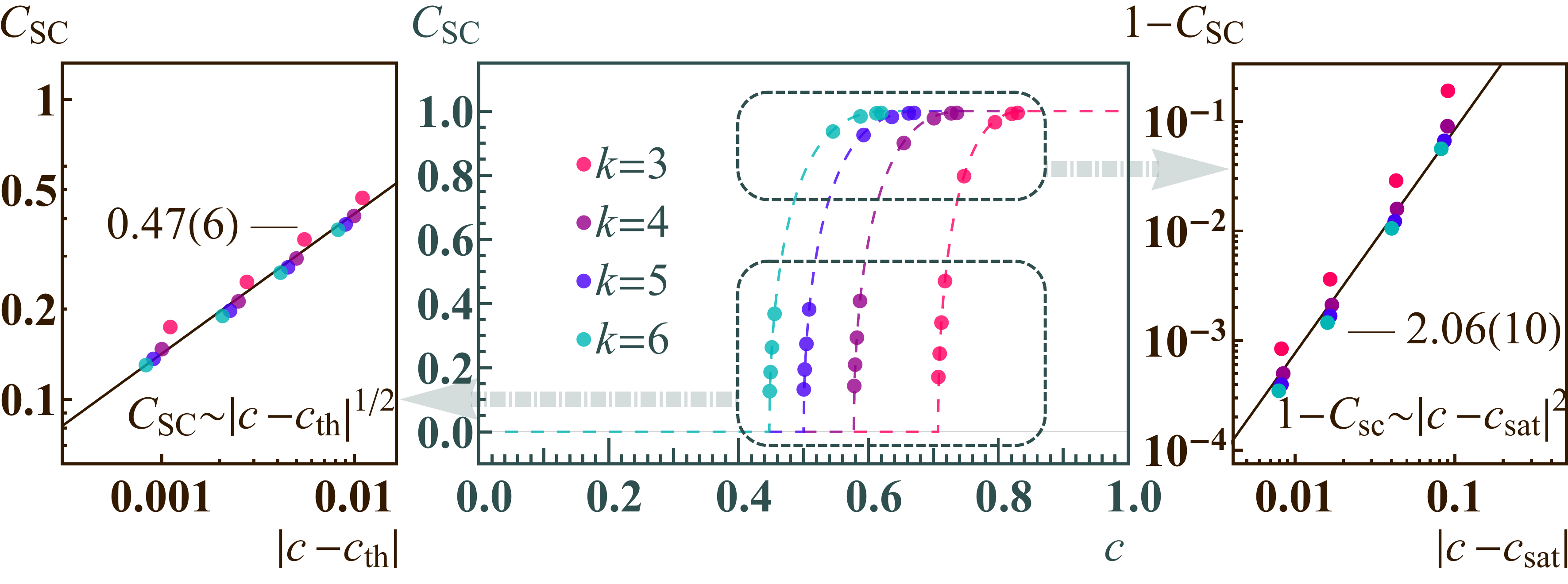}
		\vspace{-9mm}
		\subcaption{\label{fig_bethe_scaling_c3s}}
	\end{minipage}
	\hspace{22mm}
	\addtocounter{subfigure}{+4}
	\begin{minipage}[t]{56mm}
		\raggedleft
		\includegraphics[height=31mm]{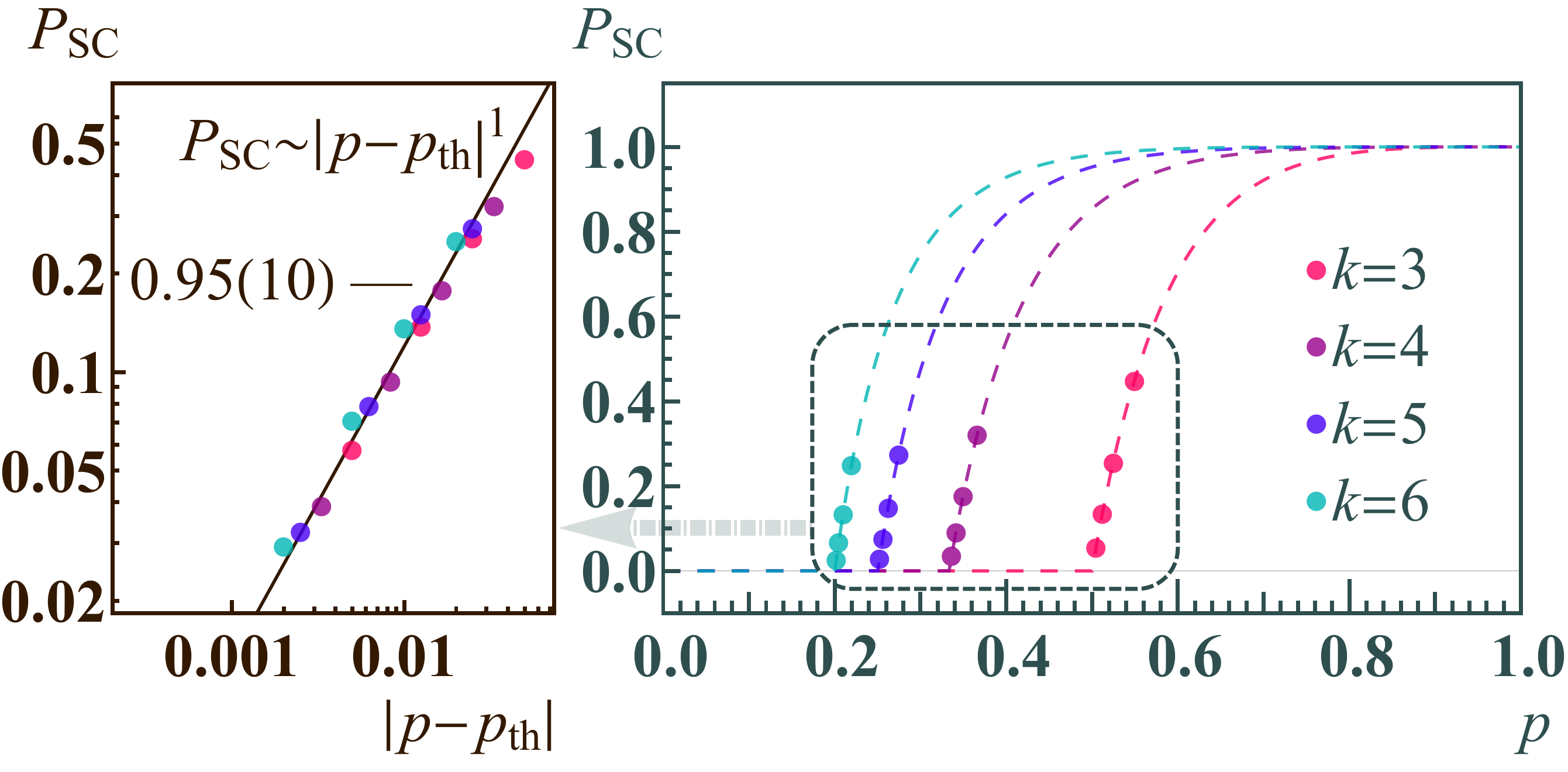}
		\vspace{-9mm}
		\subcaption{\label{fig_bethe_scaling_p3}}
	\end{minipage}		
	\hspace{100mm}

	\caption{\label{fig_bethe}
		Comparison of the universal behaviors between \subref{fig_bethe_scaling_c10}--\subref{fig_bethe_scaling_c3s}~ConPT and \subref{fig_bethe_scaling_p10}--\subref{fig_bethe_scaling_p3}~classical percolation theory for the Bethe lattice by finite-size analysis.
		\hfill\hfill}
\end{figure}

\subsection{Exact renormalization on the Bethe lattice in terms of series and parallel rules}

In the Bethe lattice, we can select an arbitrary node to be the root and then construct an exact recurrence relation between the root and the subroots at the top of the branches. The renormalization trick yields
\begin{equation}
\label{renorm-trick}
P'_{\text{SC}}=\text{para}\stackrel{k-1}{\overbrace{\left(y,y,\cdots,y\right)}},\qquad
y=\text{seri}\left(P'_{\text{SC}},p\right),\qquad
P_{\text{SC}}=\text{para}\left(P'_{\text{SC}},y\right).
\end{equation}
Classically we solve $P_{\text{SC}}$ in terms of $p$ and find that $P_{\text{SC}}=0$ holds true until $p\ge p_\text{th}$. For example, when $k=3$, $P_{\text{SC}}=1-(1-p)^3/p^3$ if $1/2\le p\le1$ and vanishes otherwise. $P_{\text{SC}}$ behaves like the order parameter, and we find $P_{\text{SC}}\sim\left|p-p_\text{th}\right|^1$ near $p_\text{th}$ for all $k$. 

For ConPT, we solve Eq.~(\ref{renorm-trick}) but using the rules for concurrence to find $c_\text{th}$ under the critical condition $C_{\text{SC}}=0$. The saturation points $c_\text{sat}$ are located using another critical condition $C_{\text{SC}}=1$. When $k=3$, $C_{\text{SC}}=\sin\{2\cos^{-1}[2^{-3/2}{(\sqrt{4+\sin ^2 2\theta} /\sin2 \theta-1)^{3/2}}]\}$ as $c_\text{th}\le c\equiv\sin2\theta\le c_\text{sat}$. By series expansion we find, for all $k$,  $C_{\text{SC}}\sim\left|c-c_{\text{th}}\right|^{1/2}$ near $c_{\text{th}}$ 
and $1-C_{\text{SC}}\sim\left|c-c_{\text{sat}}\right|^{2}$ near $c_\text{sat}$, respectively.

Furthermore, we can calculate the sponge-crossing quantities recursively for finite-size Bethe lattices (i.e.,~for finite number of layers $l$). Shown in Fig.~\ref{fig_bethe} are the results not only for ConPT but also for classical percolation theory as a comparison. Figures~\ref{fig_bethe_scaling_c11s},~\ref{fig_bethe_scaling_c21s},~and~\ref{fig_bethe_scaling_c3s} are also shown in the main paper where details of the finite-size analysis are described.

\subsection{Comparison of different strategies}

\begin{figure}[h]
	\centering
	\includegraphics[width=80mm]{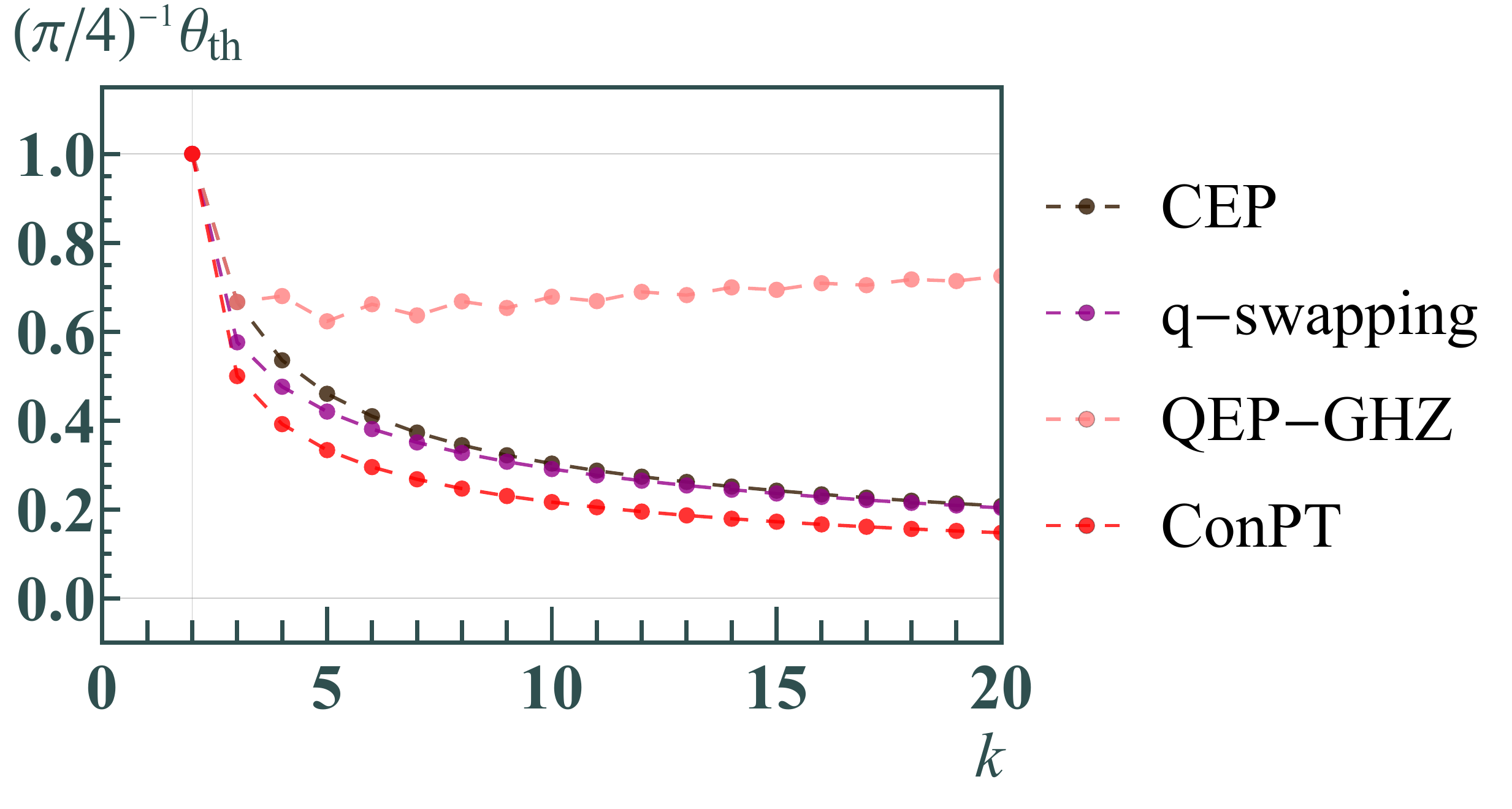}
	\caption{\label{fig_bethe-strategy}Comparison of the entanglement transmission thresholds yielded by different LOCC strategies (CEP~\cite{QEP_acl07}, $q$-swapping~\cite{QEP-q-swap_cc09}, and QEP-GHZ~\cite{QEP-GHZ_pclla10}) for the Bethe lattice of different $k$. \hfill\hfill}
\end{figure}

The Bethe lattice is a typical series-parallel network. Therefore, one can simply design an LOCC strategy using only the series and parallel rules to reach the ConPT threshold. Compared with other previously known classical-percolation-theory-based strategies (Fig.~\ref{fig_bethe-strategy}), we can see that the ConPT threshold remains the lowest for all $k$. (Note that QEP-GHZ as a special multipartite strategy actually converts the problem of CEP to a new statistical problem in terms of classical site percolation~\cite{QEP-GHZ_pclla10}, the new threshold of which, however, is the same as the classical bond percolation threshold for the Bethe lattice. Therefore QEP-GHZ shows no advantage over CEP; in fact, it performs much worse than CEP.)

When processing our LOCC strategy, none of the links is converted to a singlet, which is fundamentally different from the other three strategies; still, one can achieve nonzero entanglement transmission for infinitely long distance. The existence of such a strategy clearly relaxes the necessity of establishing singlets.

\newpage
\subsection{Diluted Bethe lattice}

\begin{figure}[h!]
	\centering
	\includegraphics[width=65mm]{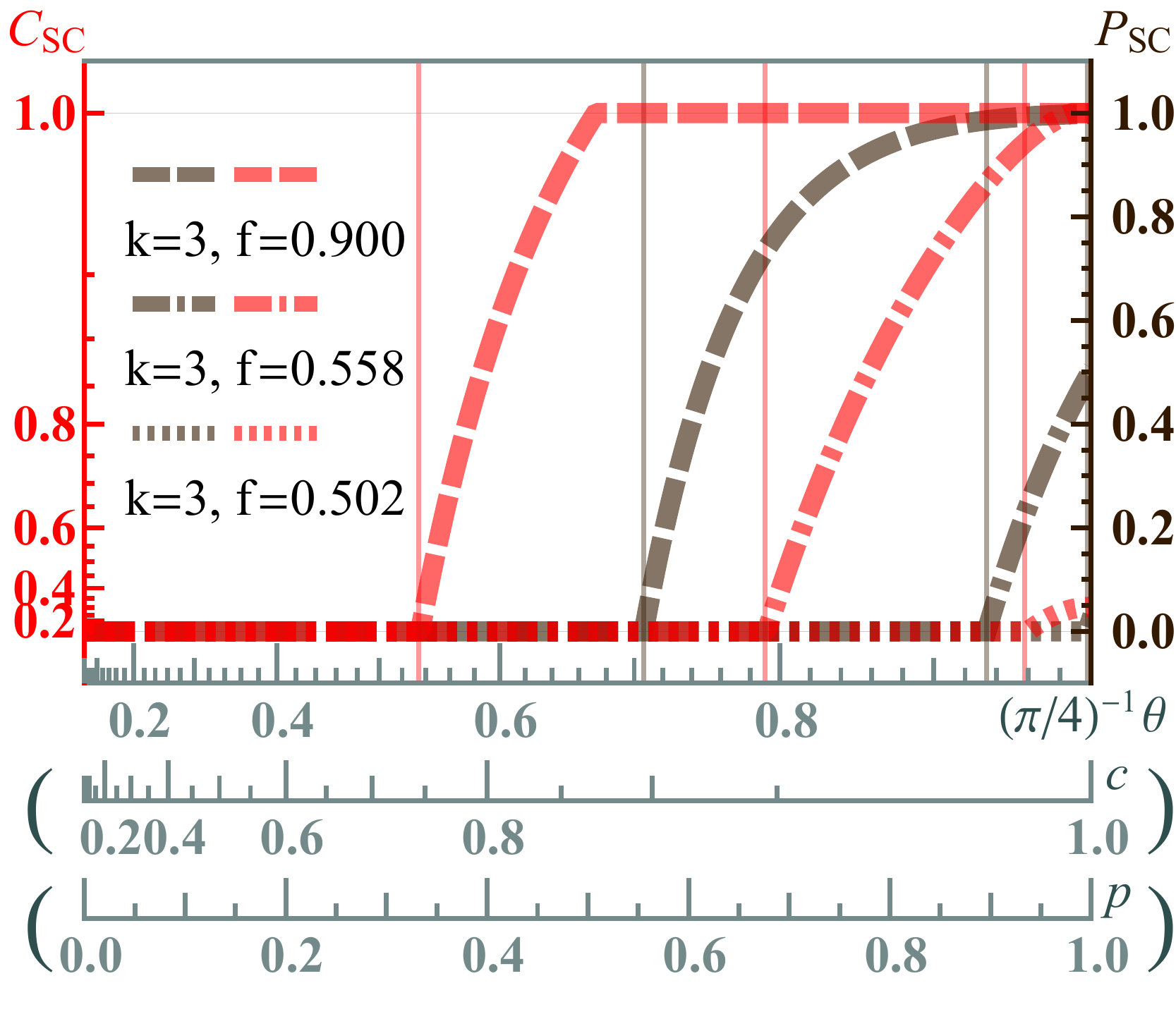}
	\caption{\label{fig_critical-curve_dilute-bethe}After removal of $1-f$ fraction of the links in the Bethe lattice, $c_{\text{th}}$ and $p_{\text{th}}$ all shift to the right. Both $P_\text{SC}$ and $C_\text{SC}$ are calculated by Eq.~(\ref{renorm-trick-dilute}) [substituted for Eq.~(\ref{renorm-trick})]. Note that unlike $P_\text{SC}$, $C_\text{SC}$ is not the ``true'' sponge-crossing concurrence unless near $c_{\text{th}}$. \hfill\hfill}
\end{figure}

Now, for a diluted Bethe lattice where $1-f$ fraction of the links are randomly removed, we are tempted to still use Eq.~(\ref{renorm-trick}) but change the first equation there by
\begin{equation}
\label{renorm-trick-dilute}
P'_{\text{SC}}=\sum_{m=0}^{k-1}\left[\binom{k-1}{m} f^m (1-f)^{k-1-m}\text{para}\stackrel{m}{\overbrace{\left(y,y,\cdots,y\right)}}\right]
\end{equation}
which counts for the average of possibilities of having different numbers of branches.
For classical percolation theory, the structural average is equal to the ensemble average, and therefore one only needs to replace $k-1$ by $f(k-1)$ to find the new percolation threshold $p_\text{th}=1/[f(k-1)]$. $P_\text{SC}$ calculated by Eq.~(\ref{renorm-trick-dilute}) [substituted for Eq.~(\ref{renorm-trick})] is the exact sponge-crossing probability between the root and the boundary on a diluted Bethe lattice (Fig.~\ref{fig_critical-curve_dilute-bethe}).

However, for ConPT, it is unknown if the structural average can be considered as equal to the ensemble average. Thus, $C_\text{SC}$ calculated by Eq.~(\ref{renorm-trick-dilute}) [substituted for Eq.~(\ref{renorm-trick})] cannot be simply explained as the ``true'' sponge-crossing concurrence between the root and the boundary. Nevertheless, one can still find $c_\text{th}=1/\sqrt{f(k-1)}$ which must be the exact threshold. This is because 
\begin{equation}
\sum_{m=0}^{k-1}\left[\binom{k-1}{m} f^m (1-f)^{k-1-m}\left(\frac{1+\sqrt{1-c^2}}{2}\right)^m\right]\simeq 1-f(k-1)c^2/4+\cdots
\end{equation}
when $c\to 0$. Therefore, near the threshold it is justified to replace $k-1$ by $f(k-1)$.

\section{Uniqueness of Threshold for Entanglement Transmission}

We have shown in the main paper that a nontrivial threshold exists for establishing the final concurrence between two infinitely distant nodes. However, one subtle question remains. ConPT focuses on how to optimize the establishment of the final average concurrence---which is different from the goal of CEP/QEP schemes~\cite{QEP_acl07}, i.e.,~how to optimize the final probability of establishing a singlet. It is unknown if these two goals are equivalent and are governed by one unique threshold in the thermodynamic limit. If they were not, then the definition of the existence of \emph{one} entanglement transmission threshold would  indeed be ambiguous---and one had to specify which entanglement transmission goal the underlying statistical theory is focusing on.

Interestingly, in response to this question, here we will show that there is one (and only) minimum threshold, if exists, for entanglement transmission between two infinitely distant nodes in any QN, that optimizes not only the final probability of establishing a singlet but also the final average concurrence between the two nodes. This is proved based on the following theorem.

\textbf{Theorem. }\emph{Let  ${P^*}\left[\mathcal{G}_{\theta}\left(n\right)\right]$ be the maximum singlet conversion probability (SCP) that is allowed in theory to be gained by LOCC between arbitrary two nodes in $\mathcal{G}_{\theta}\left(n\right)$ and ${C^*}\left[\mathcal{G}_{\theta}\left(n\right)\right]$ be the maximum average concurrence allowed between the same nodes. Then we have} 
\begin{equation}
\label{squeeze}
P^*\left[\mathcal{G}_{\theta}\left(n\right)\right]\le C^*\left[\mathcal{G}_{\theta}\left(n\right)\right]\le \sqrt{1-\left(1-P^*\left[\mathcal{G}_{\theta}\left(n\right)\right]\right)^2}.
\end{equation}
\textit{Proof. }Let  ${C^*}\left[\mathcal{G}_{\theta}\left(n\right)\right]=\sum_{k}{\omega_{k}C_{k}}$, where $C_{k}$ is the concurrence of outcome $k$ (a pure state), weighted by probability $\omega_k$ in the ensemble of outcomes. Similarly we can write ${P^*}\left[\mathcal{G}_{\theta}\left(n\right)\right]=\sum_{k'}{\omega'_{k'} P_{k'}}$.  %$\sum_{k}{\omega_{k}}=\sum_{k'}{\omega'_{k'}}=1$ is understood.
On the one hand, obtaining a pure state $k$ with concurrence $C_{k}$ is logically harder than obtaining an ensemble that has an equivalent SCP: $1-\sqrt{1-C_{k}^2}$, because the deterministic state $k$ can produce such an ensemble under Vidal's strategy~\cite{QEP-parallel-rule_v99} but not vice versa. So,
$\sum_{k}{\omega_{k}C_{k}}\le\sum_{k}{\omega_{k} \sqrt{1-(1-P_{k})^2}} \le \sqrt{1-\left(1-\sum_{k}\omega_{k}{P_{k}}\right)^2}
\le \sqrt{1-\left(1-P^*[\mathcal{G}_{\theta}\left(n\right)]\right)^2}$.
This is since $f(x)=\sqrt{1-(1-x)^2}$ is monotonically increasing and concave when $0\le x\le1$, and $\sum_{k}\omega_{k}{P_{k}}  \le\sum_{k'}{\omega'_{k'} P_{k'}}$ for ${P^*}\left[\mathcal{G}_{\theta}\left(n\right)\right]$ is the maximum SCP. On the other hand, an inequality $P^*\left[\mathcal{G}_{\theta}\left(n\right)\right]\lesssim C^*\left[\mathcal{G}_{\theta}\left(n\right)\right]$ has been shown for one-dimensional (1D) chains~\cite{QEP-detail_pcalw08}. This can be understood in a strict manner for arbitrary $\mathcal{G}_{\theta}\left(n\right)$:
for each pure state $k'$, we know that its concurrence is strictly no less than its SCP ($\sin 2\theta \ge 2\sin^2\theta$),
which thus yields $\sum_{k'}{\omega'_{k'} P_{k'}} \le \sum_{k'}{\omega'_{k'} C_{k'}} \le \sum_{k}{\omega_{k}C_{k}}$, a result of $C^*\left[\mathcal{G}_{\theta}\left(n\right)\right]$ being the maximum average concurrence.
Thus Eq.~(\ref{squeeze}) is proved by joining the results together.
\hfill\qed

\textbf{Remark. }
We can see that  $P^*\left[\mathcal{G}_{\theta}\left(n\right)\right]$ and $C^*\left[\mathcal{G}_{\theta}\left(n\right)\right]$ can only be zero (or one) simultaneously. Thus, the minimum thresholds optimizing either concurrence or SCP are equal after squeezing Eq.~(\ref{squeeze}) to the thermodynamic limit $n\rightarrow\infty$.
It is thus possible to focus on optimizing the transmission of concurrence---instead of the commonly used approach of establishing a path of singlets---to study entanglement transmission.

Note that one explicit advantage of the series and parallel rules introduced for ConPT is that they yield the maximum obtainable average concurrence for \emph{either} series [Fig.~\ref{fig_rules_a}] \emph{or} parallel topology~[Fig.~\ref{fig_rules_b}]. As a comparison, it is known that the series and parallel rules for classical percolation theory \emph{neither} maximize the SCP for the series \emph{nor} the parallel topology~\cite{QEP_acl07}. This fact has been mentioned in the main text.
However, given an arbitrary series-parallel network which contains \emph{both} series \emph{and} parallel topologies, the ConPT rules may not yield the maximum average concurrence. This is because for some series-parallel combinations, a global and nondeterministic LOCC may exist which can produce larger average concurrence than what the local series and parallel rules can yield. Thus, it remains an open question what the maximum obtainable average concurrence $C^*\left[\mathcal{G}_{\theta}\left(n\right)\right]$ is and how to determine it. As discussed, this question is equivalent to determining the maximum SCP $P^*\left[\mathcal{G}_{\theta}\left(n\right)\right]$ too when $n\rightarrow\infty$.

\end{document}